\documentclass[manuscript=article]{achemso}            		

\setkeys{acs}{usetitle=true}

\usepackage{graphicx}
\usepackage{subfigure}
\usepackage{multirow}

\usepackage{chemformula} 
\let\ce\ch
\usepackage{longtable}
\usepackage{adjustbox}

\usepackage[utf8]{inputenc} 
\usepackage{booktabs}
\usepackage{multirow}
\usepackage{lscape}
\usepackage{caption}
\usepackage{xcolor}

\title{Are Universal Potentials Ready for Alkali-Ion Battery Kinetics?}

\author{Xingyu Guo}
\email{xingyguo@cityu.edu.hk}
\affiliation{Department of Data Science, City University of Hong Kong, Hong Kong SAR, 999077, China}
\alsoaffiliation{Hong Kong Institute of AI for Science, Hong Kong SAR, 999077, China}
\author{Cheng Gui}
\affiliation{Hong Kong Institute of AI for Science, Hong Kong SAR, 999077, China}
\author{Zhenbin Wang}
\affiliation{Department of Materials Science and Engineering, City University of Hong Kong, Hong Kong SAR, 999077, China}
\alsoaffiliation{School of Energy and Environment, City University of Hong Kong, Hong Kong SAR, 999077, China}

\begin{document}
\maketitle

\begin{abstract}

Accelerating alkali-ion battery discovery requires accurate modeling of atomic-scale kinetics, yet the reliability of universal machine learning interatomic potentials (uMLIPs) in capturing these high-energy landscapes remains uncertain. Here, we systematically benchmark state-of-the-art uMLIPs, including M3GNet, CHGNet, MACE, SevenNet, GRACE, and Orb, against DFT baselines for cathodes and solid electrolytes. We find that the Orb-v3 family excels in static migration barrier predictions (MAE $\approx$ 75--111 meV), driven primarily by architectural refinements. Conversely, for dynamic transport, the GRACE model trained on the OMat24 dataset demonstrates superior fidelity in reproducing ion diffusivities and structural correlations. Our results reveal that while architectural sophistication (e.g., equivariance) is beneficial, the inclusion of high-temperature, non-equilibrium training data is the dominant driver of kinetic accuracy. These findings establish that modern uMLIPs are sufficiently robust to serve as zero-shot surrogates for high-throughput kinetic screening of next-generation energy storage materials.

\end{abstract}

\section{Introduction}

Alkali-ion batteries (AIBs) are central to the sustainable energy transition, with applications spanning portable electronics to grid-scale storage.\cite{bruce2008energy} The performance of these devices is fundamentally governed by the atomic-scale behavior of their electrode and electrolyte materials. Although thermodynamics dictates the open-circuit voltage and theoretical capacity of battery materials\cite{goodenough2013li}, it is kinetics, including ion transport within bulk phases and across interfaces\cite{banerjee2020interfaces,guo2025kinetic}, phase boundary dynamics\cite{lee2023atomic}, and electrochemical reactions\cite{tang2018probing}, that ultimately determines practical power density and operational lifetime. Understanding and predicting these kinetic phenomena is therefore essential for overcoming the rate-limiting steps that currently constrain the development of high-performance battery chemistries.

Density functional theory (DFT) has long served as the computational foundation for investigating battery kinetics at the atomic scale.\cite{ceder1997application,van2020rechargeable} However, its substantial computational cost restricts simulations to relatively small system sizes and short timescales, creating a fundamental ``scale gap" that prevents effective sampling of the complex diffusion networks governing ion transport in realistic battery materials. This limitation has driven an urgent demand for computational approaches that achieve DFT-level accuracy while enabling large-scale screening and comprehensive kinetic modeling.

Recent advances in machine learning interatomic potentials (MLIPs) provide a robust strategy to bridge this gap, offering near-quantum accuracy at a fraction of the computational cost\cite{jacobs2025practical}. Specifically, universal MLIPs (uMLIPs), trained on vast datasets spanning the periodic table, have emerged as viable general-purpose surrogates for DFT. The predictive fidelity of uMLIPs for both thermodynamic properties (e.g., formation energies)\cite{focassio2024performance, shuang2025universal} and dynamic properties (e.g., phonon spectra)\cite{loew2025universal, focassio2024performance} has seen remarkable improvement. As illustrated in Figure 1, this progress is driven by the co-evolution of model architectures and training datasets. On the architectural front, the field has transitioned from invariant graph neural networks (GNNs), such as M3GNet, to sophisticated equivariant frameworks including NequIP\cite{batzner20223}, MACE\cite{Batatia2022Design, Batatia2022mace}, and GRACE\cite{lysogorskiy2025graph}. In parallel, training datasets have evolved from the near-equilibrium structures of Materials Project Trajectories\cite{jain2013commentary,deng2023chgnet} to the high-diversity Alexandria\cite{schmidt2023machine}, and OMat datasets\cite{barroso2024open}. This synergistic integration has enabled state-of-the-art (SOTA) models to achieve exceptional accuracy, yielding minimal deviations in energies, forces, and stresses compared to DFT benchmarks.

Despite these advances, the capability of uMLIPs to predict kinetic properties remains largely unexplored. Kinetic processes inherently involve high-energy transition states, yet most uMLIPs are trained and validated predominantly on near-equilibrium or ground-state structures. Consequently, it remains unclear whether these models can reliably capture diffusion kinetics and related phenomena such as defect migration. Moreover, architectural limitations in many SOTA models, including non-conservative force predictions and discontinuities in predicted potential energy surfaces\cite{fu2022forces}, may further compromise their accuracy for kinetic property predictions. Addressing this knowledge gap is critical for developing improved uMLIPs suitable for both thermodynamic and kinetic predictions. 

In this work, we systematically benchmark the predictive capabilities of uMLIPs for modeling ion transport in battery materials using both nudged elastic band (NEB) calculations and molecular dynamics (MD) simulations. For migration barrier predictions, we find that the Orb‑v3 model achieves the lowest errors among all tested models, indicating that architectural improvements contribute more significantly to accuracy gains than expanding training data for static barriers. In contrast, for MD simulations, the GRACE model trained on OMat24 demonstrates the best overall performance, accurately reproducing mean square displacements (MSDs), ion diffusivities, and radial distribution functions (RDFs). Collectively, these results show that current state‑of‑the‑art uMLIPs enable predictive modeling of materials kinetics without system‑specific fine‑tuning. More broadly, our analysis reveals an interplay between model architecture and training data quality: while architectural sophistication is critical for energetic barrier predictions, data scale and diversity, particularly the inclusion of high‑temperature and non‑equilibrium structures, are essential for reliably capturing dynamic transport properties.

\begin{figure}[htp!]
\centering
{\includegraphics[width=\textwidth]{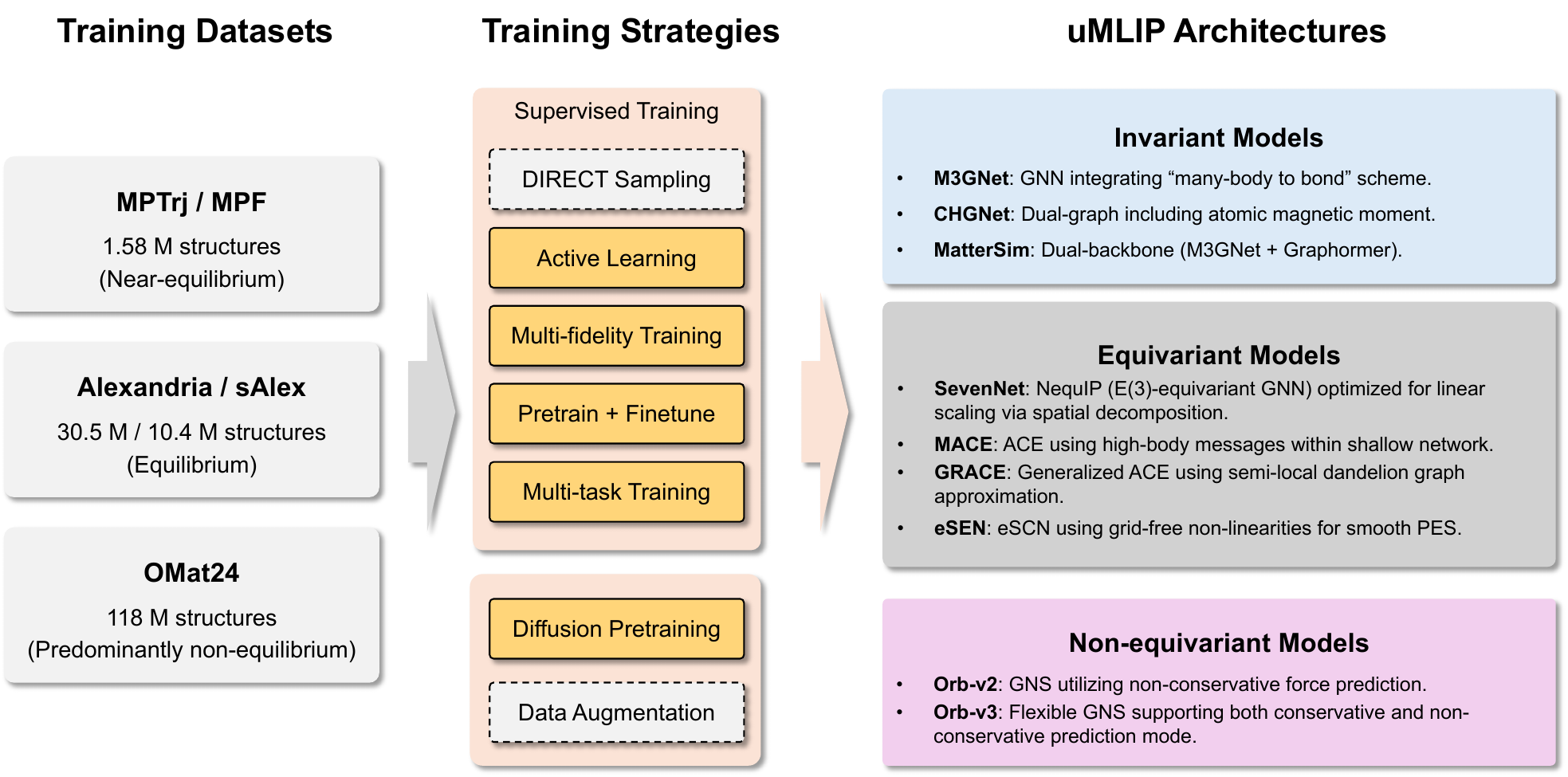}}
\caption{Overview of the uMLIP development ecosystem. The schematic illustrates the progression from training datasets, categorized by scale and thermodynamic state, through a range of training strategies to distinct model architectures. Architectures are grouped by their geometric symmetry properties (invariant, equivariant, and non‑equivariant), with representative models and defining structural features highlighted for each category.}
\label{fig:umlip_overview} 
\end{figure}

\section{Results}
\subsection{Benchmark Dataset Construction}
A diverse dataset of well-known cathode materials for both Li- and Na-ion batteries was curated to benchmark the accuracy of uMLIPs in predicting ion migration barriers. As detailed in Table 1, this dataset comprises three major electrode structural families: layered, olivine, and maricite. For layered materials, both single- and di-vacancy migration mechanisms (Figure S1) were investigated, as the latter has been reported as the dominant transport pathway due to its significantly lower energy barrier. For olivine and maricite structures, only the single-vacancy mechanism was evaluated. The dataset's chemical diversity was enhanced by incorporating various transition metals (Ni, Co, Ti, Mn, Fe) and anions (O vs. S) for both Li and Na ions within each structural framework. Additionally, to facilitate a direct comparison between alkali ions, migration barriers were calculated for each material alongside its corresponding Li- or Na-ion counterpart.

\begin{table}[H]
\renewcommand{\arraystretch}{1.5}
\footnotesize
    \centering
    \begin{tabular}{cccc}
    \hline
     Materials (A = Li or Na) &  Structure Prototype &  Space Group & Migration Mechanism\\ 
    \hline
    \ce{ATiS2}  & Layered & \textit{R}$\bar{3}$m & Single vacancy; Di-vacancy \\
    \ce{ATiS2}  & Layered & \textit{P}$\bar{3}$m1 & Single vacancy; Di-vacancy \\
    \ce{AMO2} (M = Ni, Co, Ti)  & Layered & \textit{R}$\bar{3}$m & Single vacancy; Di-vacancy \\
    \ce{AMPO4} (M = Mn, Fe, Co, Ni)  & Olivine & \textit{Pnma} & Single vacancy \\
    \ce{AMPO4} (M = Mn, Fe, Co, Ni) & Maricite & \textit{Pnma} & Single vacancy \\
    \hline
    \end{tabular}
    \caption{Selected electrode materials for Li$^+$ and Na$^+$ migration barriers evaluation.}
    \label{tab:table_cathode_mat}
\end{table}

The solid electrolyte materials listed in Table 2 were selected to provide a chemically and structurally diverse benchmark set for evaluating uMLIP performance in predicting ion diffusivity. This selection represents key material families currently under active study for all-solid-state batteries, including sulfides, argyrodites and halides. These materials encompass a wide range of bonding environments and diffusion mechanisms, with ionic conductivities spanning several orders of magnitude from highly conductive sulfides such as \ce{Li10GeP2S12} and \ce{Li20Si3P3S23Cl} to more resistive systems like \ce{Na3PS4} and \ce{Na10GeP2S12}. By incorporating various structural prototypes (e.g., LGPS, halides, NASICONs), this dataset ensures a rigorous assessment of uMLIP transferability and generalizability across multiple crystal symmetries and ion migration dimensionalities.

\begin{table}[H]
\renewcommand{\arraystretch}{1.5}
\footnotesize
\centering
\begin{tabular}{cccc}
\hline
Material Formula & Structure Prototype & Space Group & Ref\\
\hline
\ce{Li10GeP2S12} & LGPS & $P$4$_2$/$nmc$ & \citenum{kamaya2011lithium} \\
\ce{Li10SiP2S12} & LGPS & $P$4$_2$/$nmc$ & \citenum{kim2019structures} \\
\ce{Li10SnP2S12} & LGPS & $P$4$_2$/$nmc$ & \citenum{bron2013li10snp2s12}\\
\ce{Li20Si3P3S23Cl} & LGPS & $P$4$_2$/$nmc$ & \citenum{kato2016high}\\
\ce{Li3YCl6} & Halide & $P$$\bar{3}$$m$1 & \citenum{asano2018solid} \\
\ce{Li3YBr6} & Halide & $P$$\bar{3}$$m$1 & \citenum{asano2018solid}  \\
\ce{Li7P3S11} & Sulfide & $P$$\bar{1}$ & \citenum{zhou2022li7p3s11}\\
\ce{Li6PS5Cl} & Argyrodite  &  $F$$\bar{4}$3$m$ & \citenum{wang2018high}  \\
\ce{Na10GeP2S12} & LGPS & $P$4$_2$/$nmc$ & \citenum{tsuji2018preparation} \\
\ce{Na10SiP2S12} & LGPS & $P$4$_2$/$nmc$ & \citenum{richards2016design}\\
\ce{Na11Sn2PS12} & LGPS & $P$4$_2$/$nmc$ & \citenum{duchardt2018vacancy} \\
\ce{Na3PS4} & Sulfide & $P$$\bar{4}$2$_1$$c$ & \citenum{hayashi2012superionic}\\
\ce{Na3PSe4} & Sulfide &  $P$$\bar{4}$2$_1$$c$ & \citenum{bo2016computational}\\
\ce{Na3SbS4} & Sulfide &  $P$$\bar{4}$2$_1$$c$ & \citenum{banerjee2016na3sbs4} \\
\hline
\end{tabular}
\caption{Selected solid electrolyte materials for Li$^+$ and Na$^+$ kinetic properties evaluation.}
\label{tab:table_SE_mat}
\end{table}

\subsection{Reliability and stability of uMLIPs in NEB calculations}

Before evaluating the accuracy of predicted migration barriers, we first assessed the fundamental ability of uMLIPs to complete NEB calculations without computational or physical breakdown. We observed that calculations failed in two primary ways: (i) computational divergence (Type I), where atomic forces became numerically unstable during the structural relaxation process; and (ii) unphysical barriers (Type II), where the predicted transition state energy was lower than that of the initial or final states. To ensure a fair comparison of barrier accuracy, these failed calculations were excluded from the error analysis.

\begin{figure}[htp!]
\centering
{\includegraphics[width=\textwidth]{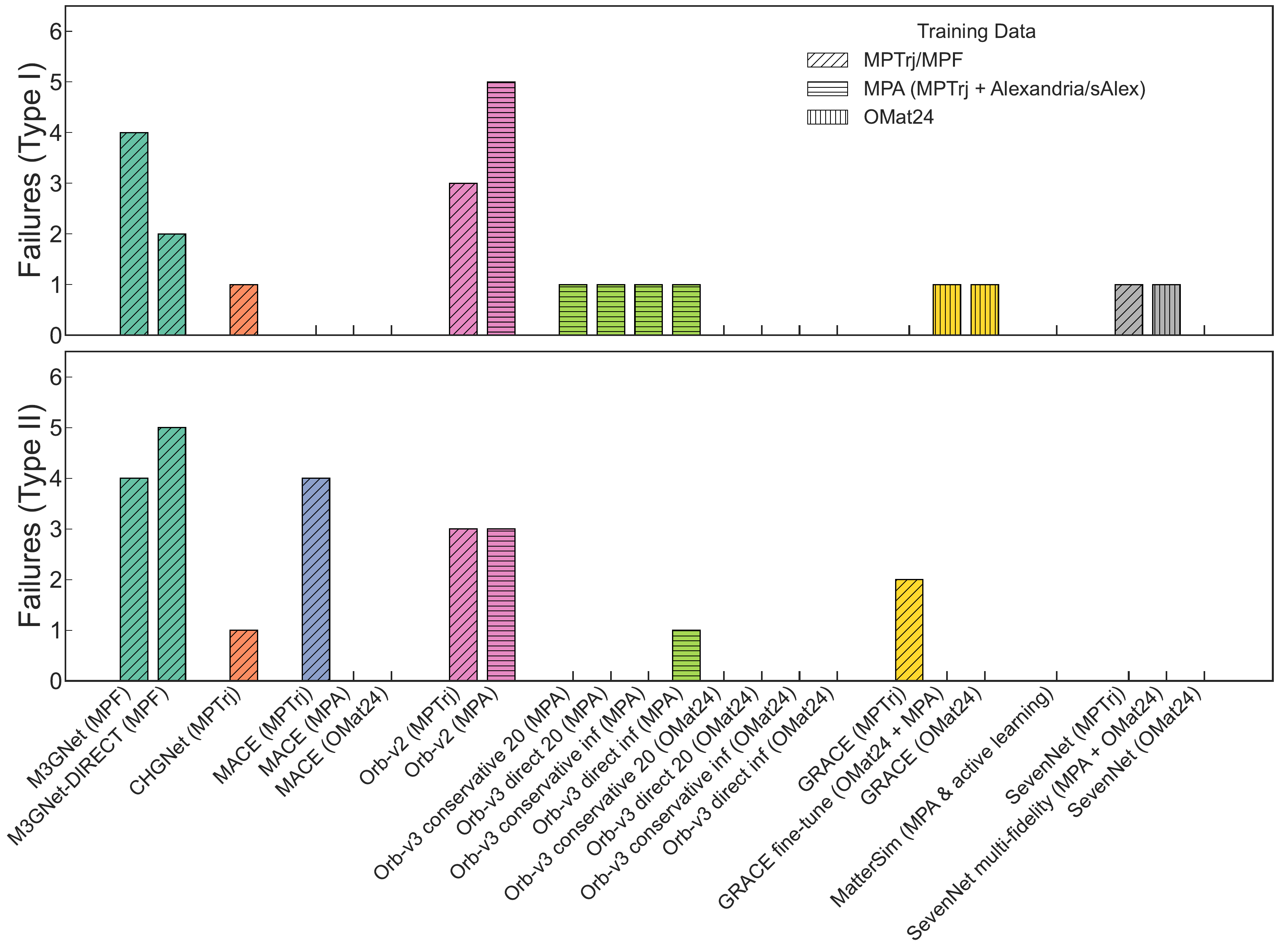}}
\caption{Number of NEB calculation failures across various uMLIP models. Failures are categorized as (top) Type I: computational divergence and (bottom) Type II: unphysical barriers (where $E_{\mathrm{TS}} < E_{\mathrm{initial/final}}$).}
\label{fig:neb_failed} 
\end{figure}

Figure \ref{fig:neb_failed} summarizes the number of failures encountered by each uMLIP across the 44 distinct migration cases tested. A clear trend emerges with respect to training data: models trained exclusively on the MPF or MPTrj datasets, which are dominated by near-equilibrium structures, consistently exhibited higher failure rates. Specifically, M3GNet variants trained only on MPF recorded the highest number of failures (8 for M3GNet and 7 for M3GNet-DIRECT). Similarly, CHGNet and Orb-v2 (MPTrj-only) showed multiple instances of instability. In contrast, models trained on more diverse datasets that explicitly include non-equilibrium structures demonstrated markedly greater robustness. For instance, models from the MACE, GRACE, SevenNet and conservative Orb-v3 families, trained on MPA and OMat24 data, exhibited one to no failures. These results underscore the critical importance of incorporating out-of-equilibrium structures into training datasets to ensure a smooth potential energy surface (PES) capable of handling high-energy transition states and distorted atomic environments.

It is also worth noting the impact of model architecture, particularly within the Orb family. The Orb-v2 model exhibited the highest frequency of Type I failures (5 cases), even when trained on the more diverse MPA dataset. This suggests that the PES learned by Orb-v2 can become excessively rugged or discontinuous under non-equilibrium conditions. By contrast, the Orb-v3 family demonstrated significant improvements. Across all conservative and direct Orb-v3 models trained on the same data, only one Type I failure was observed, with one additional unphysical barrier in the direct model with infinite neighbors. These findings highlight the benefits of architectural refinements in Orb-v3, which appear to reduce PES discontinuities. Furthermore, models grounded in the atomic cluster expansion (ACE) formalism, such as MACE and GRACE, exhibited exceptional robustness, likely due to their strong mathematical foundations. Similarly, MatterSim trained via active learning on a vast dataset achieved excellent reliability with zero recorded failures.

\subsection{Accuracy of uMLIPs in predicting migration barriers}

\begin{figure}[htp!]
\centering
{\includegraphics[width=\textwidth]{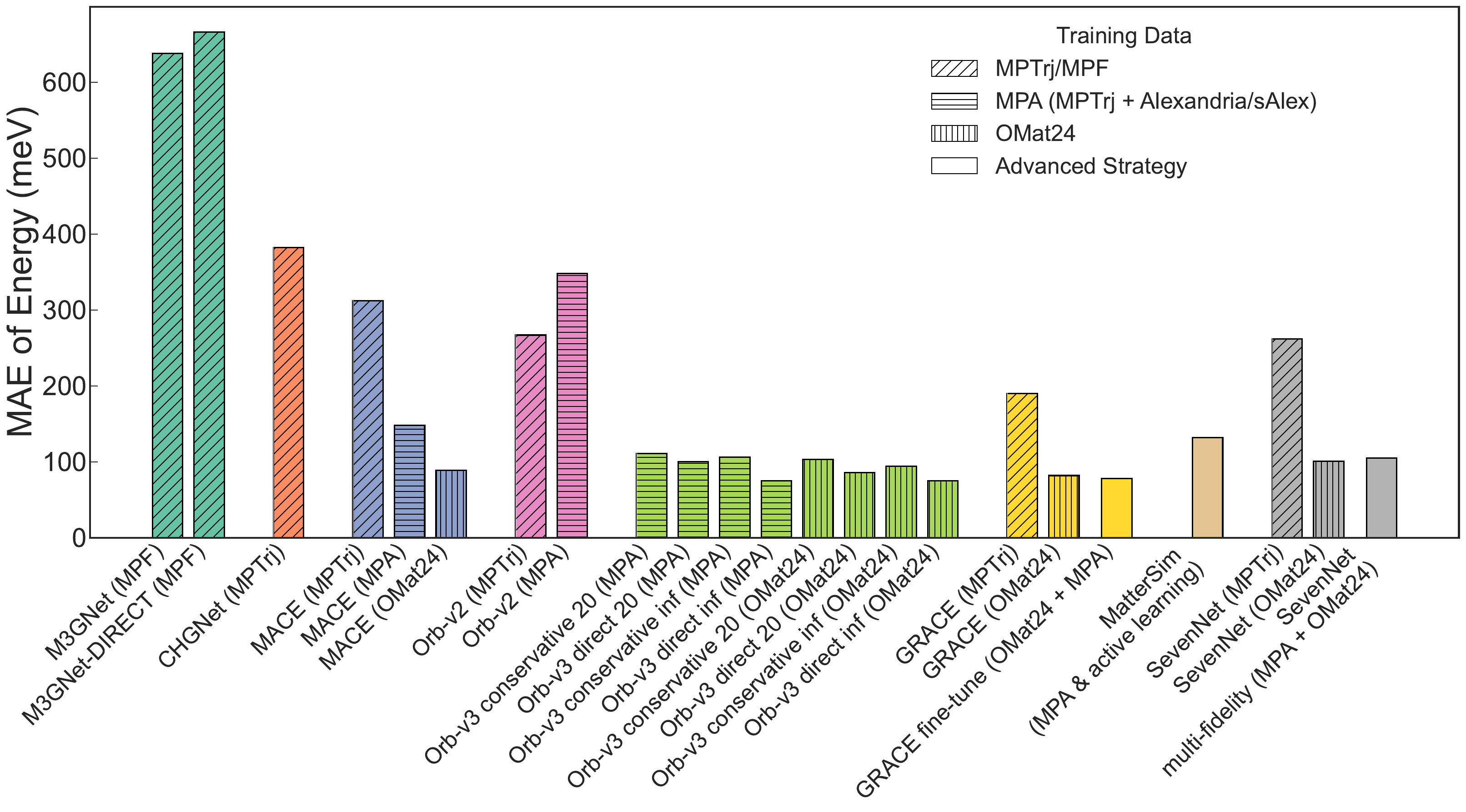}}
\caption{Mean absolute error (MAE) of energy barriers predicted by various uMLIP models relative to DFT benchmarks. The models are grouped by architecture, with bar patterns indicating the specific training dataset used.}
\label{fig:energy_barrier} 
\end{figure}

Figure \ref{fig:energy_barrier} presents the mean absolute errors (MAEs) for \ce{Li+}/\ce{Na+} migration barriers predicted by each uMLIP against DFT calculations. Among models trained exclusively on the MPF/MPTrj datasets, M3GNet exhibits the poorest performance, with the original version recording an MAE of 638 meV and the DIRECT variant yielding a slightly higher value of 666 meV. CHGNet achieves a significantly lower MAE of 382 meV, likely due to its explicit prediction of atomic magnetic moments, which serve as a proxy for charge states important for describing transition metals in cathode materials\cite{deng2023chgnet}. Further improvements are seen with equivariant models: MACE records an MAE of 312 meV, while GRACE shows a further reduction to 190 meV. SevenNet (262 meV) and Orb-v2 (267 meV) also clearly outperform the earlier invariant GNN-based models.

Significant performance gains are observed in the evolution from Orb-v2 to Orb-v3 when trained on MPA datasets. Orb-v2 exhibits a high MAE of 348 meV, whereas all Orb-v3 models achieve substantially lower values ranging from 75 to 111 meV. These values fall within the standard DFT uncertainty of $\sim$0.1 eV \cite{ganesh2014binding, urban2016computational, norskov2009towards}, suggesting that Orb-v3 reaches the practical accuracy limit of its underlying training data. This improvement stems from architectural updates, such as smoother embeddings and improved force stability\cite{rhodes2025orbv3}. Notably, within the Orb-v3 family, the non-conservative ``direct" models yield MAEs of 75–100 meV, slightly outperforming their conservative counterparts (106–111 meV). Although direct models have known limitations such as energy non-conservation that may cause instabilities in structure optimization\cite{bigi2024dark}, our results indicate these issues do not significantly impact predicted ion migration barriers for Orb-v3. Additionally, removing the fixed neighbor limit (``infinite neighbors") improves accuracy by 25 meV for direct models and 5 meV for conservative models.

Dataset diversity plays a critical role in model performance. Models trained exclusively on MPTrj generally exhibit poor performance, consistently underestimating ion migration barriers as illustrated in Figure 4. For the medium-sized MACE model, augmenting MPTrj with the sAlex dataset reduces the MAE from 312 meV to 148 meV. Further incorporating the OMat24 dataset, which explicitly includes non-equilibrium structures, decreases the MAE to 89 meV. Similar trends are observed for GRACE and SevenNet, where switching from MPTrj to OMat24 reduces errors by 108 meV and 161 meV, respectively. However, diminishing returns emerge for the high-performing Orb-v3 architectures; adding the OMat24 dataset yields only a marginal improvement ($\sim$10 meV) or no improvement at all.

Finally, we examined the impact of advanced training methodologies. MatterSim, developed via an active learning pipeline, shows a dramatic improvement over the passive M3GNet base model, achieving an MAE of 132 meV. In contrast, other sophisticated strategies yielded minimal gains. The multi-fidelity approach used for SevenNet-MF-ompa resulted in an MAE 4 meV higher than the OMat24-only version. Similarly, fine-tuning the OMat24-pre-trained GRACE-OAM models on sAlex and MPTrj provided only a negligible 4 meV reduction in MAE relative to the base model.

\begin{figure}[htp!]
\centering
{\includegraphics[width=\textwidth]{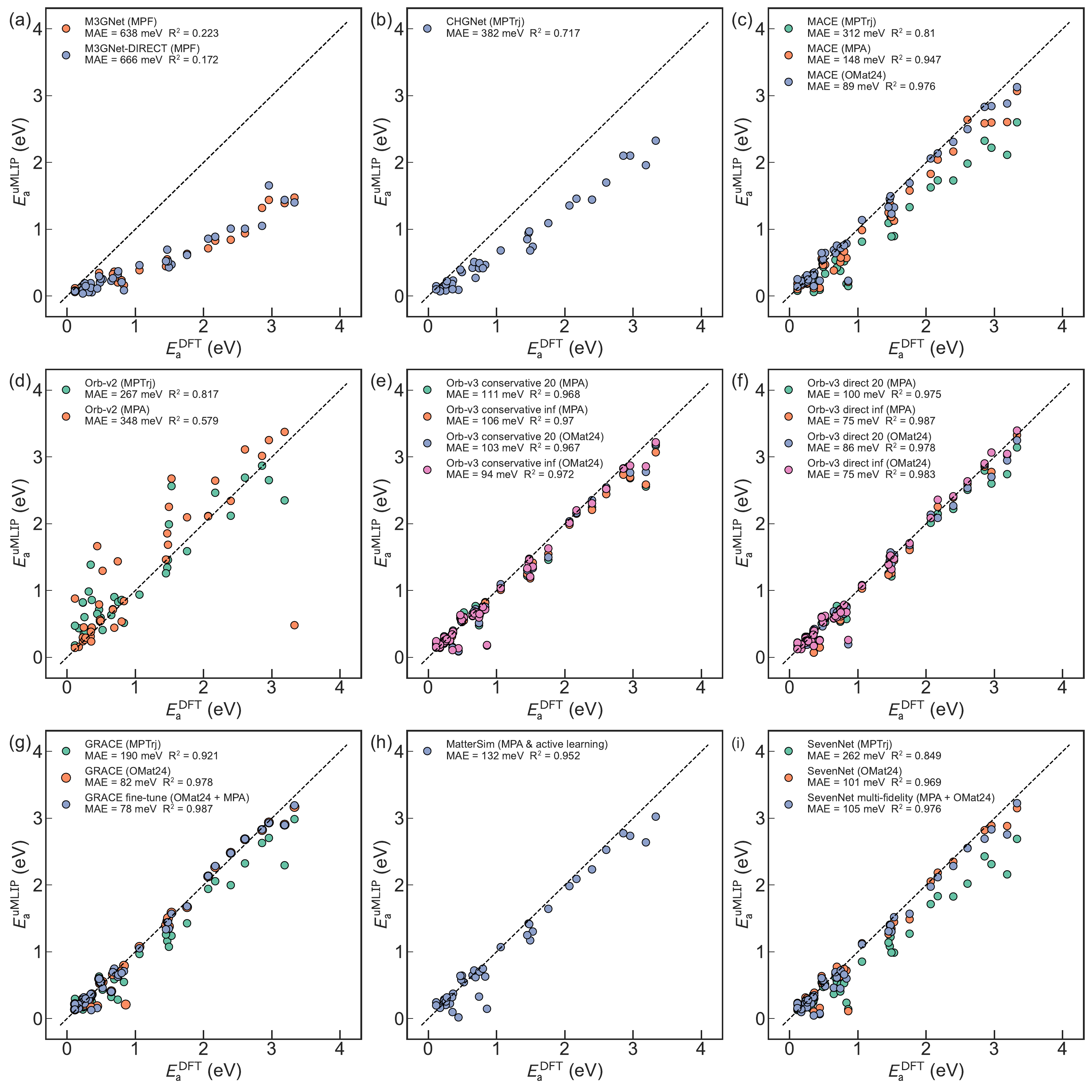}}
\caption{Comparison of ion migration barriers predicted by uMLIPs, $E_{\mathrm{a}}^{\mathrm{uMLIP}}$ with reference DFT values,  $E_{\mathrm{a}}^{\mathrm{DFT}}$. Panels (a–i) show results for different uMLIP architectures and training data, including invariant, equivariant, and non‑equivariant models trained on distinct datasets. Each point corresponds to a unique ion migration pathway obtained from NEB calculations. The dashed line indicates perfect agreement between uMLIP and DFT predictions. Mean absolute errors (MAEs) and coefficients of determination (R$^2$) are reported in each panel.}
\label{fig:energy_barrier_comparision} 
\end{figure}

\subsection{Accuracy of uMLIPs in predicting ion diffusion}
\begin{figure}[htp!]
\centering
{\includegraphics[width=\textwidth]{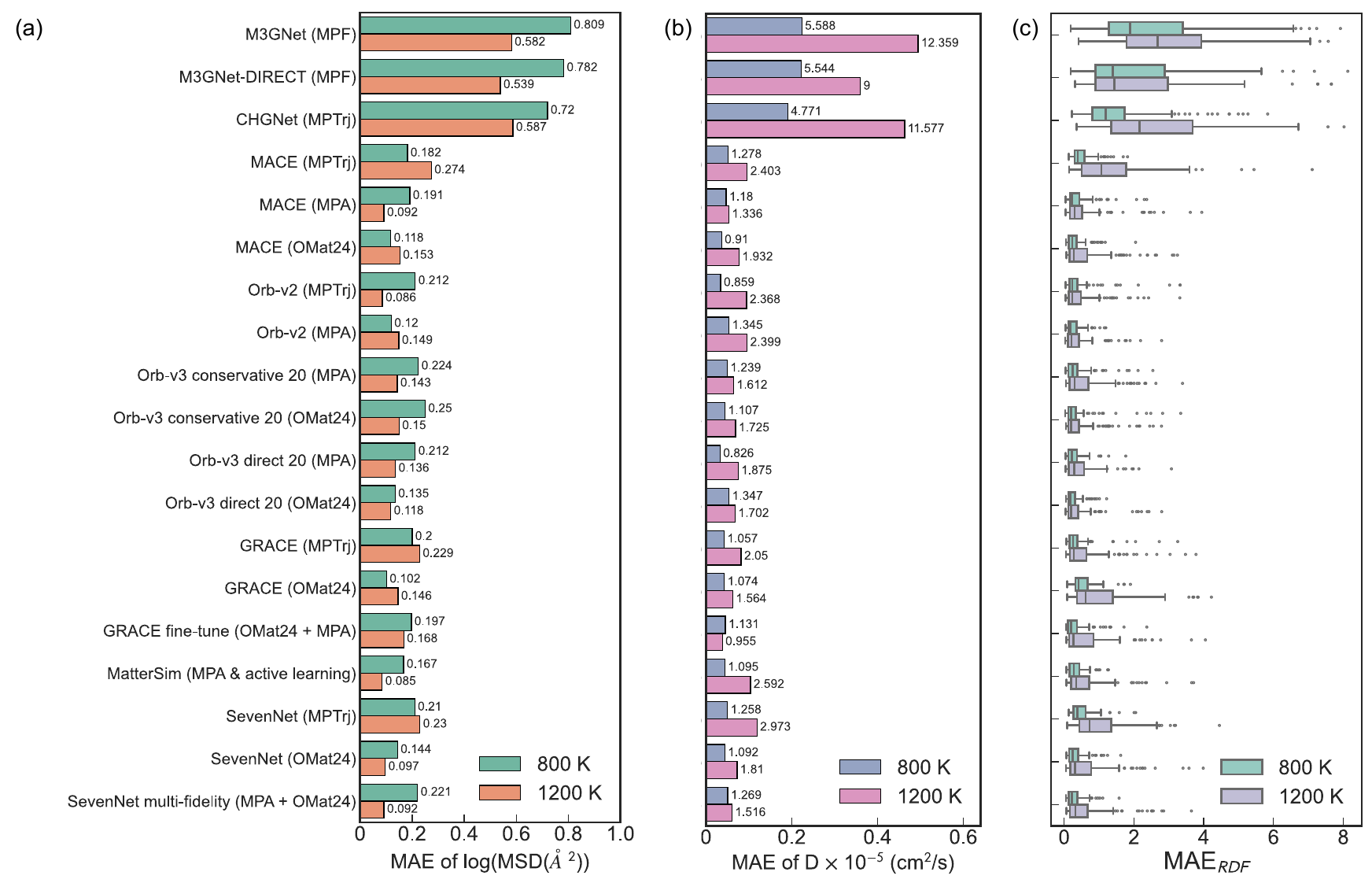}}
\caption{Performance of uMLIP molecular dynamics (MD) simulations compared with ab initio molecular dynamics (AIMD) in predicting ion transport properties. (a) Mean absolute error (MAE) of log(MSD(\r{A}$^2$)) of \ce{Li+}/\ce{Na+} ions over 50 ps of MD simulations at 800 K and 1200 K;(b) MAE of the calculated \ce{Li+}/\ce{Na+} diffusivity (10$^{-3}$ cm$^2$/s) at 800 K and 1200 K. (c) MAE of radial distribution functions (RDF), averaged over MD trajectories between 30 and 50 ps at 800 K and 1200 K.}
\label{fig:md_results} 
\end{figure}

Figure \ref{fig:md_results} presents the predictive accuracy of uMLIPs relative to AIMD calculations for three key metrics characterizing ion transport: (i) MSD of \ce{Li+}/\ce{Na+} ions, (ii) diffusion coefficients at 800 K and 1200 K, and (iii) RDFs computed from MD trajectories.

M3GNet, M3GNet-DIRECT, and CHGNet exhibit the largest errors across all three metrics. For MSD predictions (Figure \ref{fig:md_results}a), these models show MAE values of approximately 0.7--0.8 log(\r{A}$^2$) at 800 K and 0.5--0.6 log(\r{A}$^2$) at 1200 K, systematically overestimating ion mobility (Figure S2) compared with AIMD. Consistent diffusivity prediction errors appear in Figure \ref{fig:md_results}b, with MAE values of $\sim$ 5$\times$ 10$^{-5}$ cm$^2$/s at 800 K and 9--12 $\times$ 10$^{-5}$ cm$^2$/s at 1200 K. The structural accuracy metrics (Figure \ref{fig:md_results}c) reveal the underlying cause: these models display median MAE$_{RDF}$ values exceeding 1.5 with numerous outliers reaching 7--8, indicating significant inaccuracies in coordination environments and interatomic distances that directly determine energy landscapes and ion transport.

In contrast, models with equivariant features or trained on more diverse datasets exhibit substantially superior performance. These models demonstrate log(MSD) MAE values below 0.3 log(\r{A}$^2$), corresponding to prediction errors within a factor of 2. GRACE model trained on OMat24 achieves the lowest MAE of 0.102 log(\r{A}$^2$) at 800 K, while MatterSim exhibits the lowest MAE of 0.085 log(\r{A}$^2$) at 1200 K. Similar trends appear for diffusivity predictions, with MAE values below 4 $\times$ 10$^{-5}$ cm$^2$/s at both temperatures. These models also achieve median MAE$_{RDF}$ values below 0.8, with GRACE (pretrained on OMat24 $\rightarrow$ fine-tuned on MPA) and Orb-v3 variants showing the best structural accuracy (median MAE$_{RDF}$ $\approx$ 0.3–0.4).

Most models display imbalanced prediction accuracy between temperatures, potentially introducing significant errors when deriving activation energies through Arrhenius analysis. M3GNet, M3GNet-DIRECT, and CHGNet exhibit the most severe temperature-dependent bias ($\sim$ 0.2 log(\r{A}$^2$)) difference), translating to approximately 0.12–0.15 eV error in predicted activation energy. Models trained on OMat24 show more balanced performance across temperatures, likely due to more diverse temperature-dependent structures in that training dataset. GRACE (pretrained on OMat24 $\rightarrow$ fine-tuned on MPA) achieves the best overall performance, with MAE values of approximately 10$^{-5}$ cm$^2$/s at both temperatures, suggesting that combining diverse pretraining data with targeted fine-tuning provides an effective strategy for robust ionic transport predictions.

Figure S4 presents activation energy ($E_a$) correlations between uMLIP-based MD and AIMD calculations. Most models achieve MAE values within 0.10–0.20 eV, with MatterSim exhibiting the lowest MAE of 0.091 eV. However, all models display poor or negative R$^2$ values, indicating an inability to correctly rank materials by diffusivity, a limitation that could compromise reliability for high-throughput screening applications.

It is noteworthy that, although most uMLIPs demonstrate high accuracy in predicting MSD and diffusivities, all models exhibit relatively high MAE in reproducing the RDFs from AIMD simulations compared with results reported by Fu et al.\cite{fu2022forces} for LiPS, where a set of system-specific MLIPs achieved an RDF MAE of 0.04. These findings indicate potentially substantial discrepancies in structural feature analysis from uMLIPs MD simulations, particularly at elevated temperatures.

\section{Discussion}

Our comprehensive benchmarking of \ce{Li+}/\ce{Na+} transport demonstrates that SOTA uMLIPs have achieved the maturity required to accurately model kinetic properties. These models effectively capture the PES needed for large-scale kinetic screening, often yielding diffusion coefficients and migration barriers comparable to DFT. However, our analysis highlights two critical factors governing this performance: the synergistic relationship between model architecture and training data, and the decoupling of force prediction accuracy from overall kinetic barrier reliability.

\subsubsection{Interplay of Model Architecture and Data Quality}

The accuracy of uMLIPs in predicting ionic transport is determined by a critical interplay between architectural sophistication and data quality.

First, architectural advancements, specifically the inclusion of equivariant features, have clearly elevated the baseline for kinetic predictions. Models utilizing rotationally equivariant representations (e.g., MACE, SevenNet, and GRACE) consistently outperform invariant baselines when trained on comparable datasets. The superior performance of these architectures in predicting migration barriers and diffusion properties can be attributed to their more efficient use of training data and better generalization across rotational frames, which stabilizes dynamics simulations.

However, our results indicate that data scale and diversity are the dominant factors, capable of overriding architectural limitations. This is best exemplified by the comparison between M3GNet and MatterSim. Despite sharing an identical graph-based backbone, MatterSim, trained on $\sim$17 million structures via active learning, significantly outperforms M3GNet across all metrics. Remarkably, MatterSim even surpasses equivariant models trained on the smaller MPTrj dataset (including MACE, GRACE, and SevenNet). This suggests that while equivariant architectures are data-efficient, sufficiently large and diverse training datasets can compensate for the lack of equivariance.

Furthermore, the domain coverage of training data is paramount. As shown in Figure S5, models trained on datasets lacking sufficient thermal disorder (e.g., MPTraj) exhibit error amplification at high temperatures. In contrast, models trained on datasets explicitly covering high-temperature regimes (e.g., OMat24) maintain stability throughout the simulated temperature range. Consequently, for kinetic studies where MD simulations explore vast phase spaces, the inclusion of temperature-dependent structural diversity in the training set proves as critical as the choice of model architecture.

\subsubsection{Robustness of Barrier Prediction Despite Force Deviations}
A key finding of this work is the surprising robustness of uMLIP-predicted kinetic properties despite observable inaccuracies in force predictions at transition state.

Our benchmarking reveals a notable disconnect: while calculated migration barriers and ionic diffusion coefficients show excellent agreement with DFT, the predicted forces on DFT-optimized transition state structures exhibit appreciable deviations (Figure S6). Specifically, although the potential energy at saddle point is predicted with high fidelity (MAE within expected bounds; $R^2>$0.99), the force vectors, particularly components normal to the minimum energy path, often diverge from DFT reference values.

This observation carries two important implications. First, it suggests that accurate macroscopic kinetic predictions do not necessarily require perfect microscopic force reproduction at saddle point. Provided that the uMLIP accurately captures the energy topology and barrier height, deviations in forces at non-equilibrium transition state structures appear to have limited impact on derived diffusion coefficients. Second, it highlights a remaining challenge: while current uMLPs have effectively mapped the energy landscape of equilibrium structures, the force fields of high-energy, unstable configurations (such as those encountered during ion hopping) remain difficult to predict precisely.

Consequently, although current SOTA uMLIPs are highly effective for screening migration barriers and performing diffusion simulations, future development should target the reduction of force residuals in transition state regions to further enhance the accuracy of reaction pathway optimizations and kinetic prefactor predictions.



\section{Methods}

\subsection{Universal Interatomic Potentials}

\subsubsection{Models}
We evaluated a suite of SOTA uMLIPs and a brief overview of their architectures is provided below, with further details available in the original cited works.

\textbf{M3GNet (Materials 3-body Graph Network)} is a pioneering GNN-based uMLIP, building upon its predecessor MEGNet\cite{chen2019graph}. A distinguishing feature of M3GNet is its explicit incorporation of three-body interactions via a ``many-body to bond" mechanism that leverages angular information to update bond features. Furthermore, the model employs a continuously differentiable basis set expansion for bond distances, ensuring smooth predictions of energies and forces across the potential energy surface. Two M3GNet variants are utilized in this work: M3GNet-MP-2021.2.8-PES (trained on the MPF dataset)\cite{chen2022universal} and M3GNet-MP-2021.2.8-DIRECT-PES (trained on the MPF dataset using DIRECT sampling techniques)\cite{qi2024robust}.

\textbf{CHGNet (Crystal Hamiltonian Graph Network)} is a GNN-based model that employs a dual graph representation, comprising an atom graph and a bond angle graph, to incorporate three-body information. Its key feature is the explicit inclusion of atomic magnetic moments, which serve as a proxy for the charge state of atoms. By predicting magnetic moments, CHGNet can differentiate between different valence states of an element (e.g., \ce{Mn^{2+}}, \ce{Mn^{3+}}, and \ce{Mn^{4+}}). This enables the modeling of charge-dependent phenomena, particularly in cathode materials for alkali-ion batteries.\cite{deng2023chgnet} The model used in this work is CHGNet-MPtrj-2024.2.13-11M-PES, trained on the MPtrj dataset.

\textbf{MatterSim} employs a dual-backbone architecture, using M3GNet for fast inference and the more complex, transformer-based Graphormer for tasks requiring higher accuracy. It was first trained through passive learning on existing public databases, including MPF\cite{chen2022universal}, MPTrj\cite{deng2023chgnet}, and Alexandria, as well as on structures generated through Random Structure Search (RSS)\cite{pickard2011ab} and MatterGen\cite{zeni2025generative}. Subsequently, active learning with an ``off-equilibrium explorer" was employed to incorporate non-equilibrium structures generated under a wide range of temperatures (0–5000 K) and pressures (0–1000 GPa), resulting in a large, custom database of approximately 17 million atomic structures. The model used in this work is mattersim-v1.0.0-5M with an M3GNet backbone\cite{yang2024mattersim}.

\textbf{SevenNet} (Scalable EquiVariance-Enabled Neural NETwork) is an E(3)-equivariant GNN MLIP based on the NequIP\cite{batzner20223} architecture, where internal features are geometric tensors and complex many-body interactions are learned implicitly through multiple layers (typically 4–6) of message passing. Its defining characteristic is an efficient parallelization scheme based on a modified spatial decomposition that exchanges not only atomic positions but also intermediate node features and their gradients through a series of forward and reverse communication steps.\cite{park_scalable_2024,kim_sevennet_mf_2024} The models used in this work include SevenNet-0 (trained on MPTrj dataset only), SevenNet-omat (trained on the OMat24 dataset), and SevenNet-MF-ompa (trained simultaneously on MPTrj, sAlex, and OMat24 datasets using multi-fidelity learning).

\textbf{MACE} is an E(3)-equivariant message passing neural network (MPNN) based on the multi-Atomic Cluster Expansion (ACE) framework that incorporates high body-order (i.e., four-body or higher) messages within a shallow architecture. The model efficiently constructs these complex messages by taking tensor products of simpler two-body features, enabling high expressiveness with only two message-passing iterations. This design results in a smaller receptive field, making MACE significantly faster and more parallelizable than previous SOTA architectures. The models evaluated in this study include MACE-MP-0b3 (trained on the MPTrj dataset), MACE-MPA-0 (trained on MPTrj and sAlex datasets), and MACE-OMAT-0 (trained on the OMat24 dataset).\cite{Batatia2022Design, Batatia2022mace}

\textbf{GRACE (Graph Atomic Cluster Expansion)} is a complete and unified mathematical framework that generalizes local ACE by incorporating graph basis functions. Its architecture moves beyond the simple ``star" graphs of local ACE to include all admissible ``tree" graphs, which are more sensitive to complex geometries and can naturally describe semi-local interactions. The framework employs the ``dandelion approximation", a strategy mathematically equivalent to modern message-passing schemes, to enable highly efficient, recursive evaluation of global many-body interactions while maintaining high accuracy.\cite{lysogorskiy2025graph} Three variants are utilized in this work: GRACE-2L-MP-r6 (a two-layer GRACE model with a cutoff of 6 \AA, trained on MPTrj dataset), GRACE-2L-OMAT-L-base (a two-layer semi-local base model trained on the OMat24 dataset), and GRACE-2L-OAM-L (a two-layer semi-local model first trained on OMat24 and then fine-tuned on a combination of sAlex and MPTrj datasets).

\textbf{Orb-v2} utilizes a non-equivariant graph network simulator backbone to achieve high computational efficiency. Distinct from other uMLIP models, it functions as a non-conservative model that directly predicts forces and stresses rather than deriving them from total energy gradients. Rather than enforcing equivariance architecturally, the model learns rotational invariances through data augmentation during a two-phase training process: an initial denoising diffusion pretraining on unlabeled ground-state structures, followed by supervised fine-tuning to predict energies, forces, and stresses. The graph construction uses a 10 \AA\ radius cutoff with a maximum of 20 neighbors. The models used in this study are Orb-v2 (trained on the MPTrj and Alexandria datasets) and Orb-mptraj-only-v2 (trained exclusively on the MPTrj dataset)\cite{neumann2024orb}.

\textbf{Orb-v3} represents a significant evolution of the Orb framework, introducing a flexible family of models that span the Pareto frontier of accuracy, speed, and memory efficiency. While Orb-v2 was exclusively non-conservative, Orb-v3 offers both conservative (forces as energy gradients) and highly efficient non-conservative (direct) models. The direct models incorporate post-prediction corrections, including mean-subtraction to enforce zero net force and Lagrangian minimization to eliminate net torque in non-periodic systems. The conservative models employ the equigrad methodology to enhance the physical fidelity of the non-equivariant architecture. Orb-v3 also utilizes smoother edge embeddings and provides model variants without neighbor count limitations to avoid discontinuities in the PES. Eight distinct model configurations are employed: Orb-v3-direct-20-mpa, Orb-v3-direct-inf-mpa, Orb-v3-conservative-20-mpa, Orb-v3-conservative-inf-mpa, Orb-v3-direct-20-omat, Orb-v3-direct-inf-omat, Orb-v3-conservative-20-omat, and Orb-v3-conservative-inf-omat\cite{rhodes2025orbv3}.

For clarity, we label the uMLIP models according to their architectures and training datasets. A complete mapping between original model names and the notation adopted in this work is provided in Table S1. Note that the eSEN and UMA models from Meta AI were not studied due to regional access restrictions\cite{umawebsite}.

\subsubsection{Training Datasets} 
The datasets used to train these universal interatomic potentials are a key factor for understanding the capability of each model. A brief overview is provided as follows.

\textbf{MPTrj}: This dataset consists of approximately 1.58 million snapshots extracted from structural relaxation and static calculations within the Materials Project database\cite{jain2013commentary}. These calculations were performed using the PBE/PBE+U functionals and cover roughly 146,000 near-equilibrium inorganic materials spanning 89 elements. The M3GNet models were trained on the MPF dataset, which consists of 187,687 snapshots of relaxation trajectories for a smaller subset of 62,783 compounds available in the Materials Project as of February 8, 2021\cite{chen2022universal}.

\textbf{Alexandria}: This dataset initially comprised $\sim$1.89 million materials collected from three primary sources: the Automatic FLOW for Materials Discovery (AFLOW)\cite{curtarolo2012aflow}, the Materials Project\cite{jain2013commentary}, and an in-group dataset from Schmidt et al.\cite{schmidt2021crystal}. It was subsequently expanded to $\sim$30.5 million structures through elemental permutation and structural rearrangement. All calculations were performed using  parameters consistent with the Materials Project, and only equilibrium structures from structure relaxation were retained.\cite{schmidt2023machine} A refined subset of Alexandria, \textbf{sAlex}, was subsequently curated to include 10.4 million structures. This curation process involved removing structures present in the WBM dataset\cite{wang2021predicting}, discarding unstable configurations (defined by energies $>$0 eV/atom, force norms $>$50 eV/\AA, or stress $>$80 GPa), and eliminating redundant configurations that differed by less than 10 meV/atom from previously sampled structures.\cite{barroso2024open}

In practice, uMLIP models are often trained on a combined dataset of MPTrj and Alexandria, hereafter referred to as the MPA dataset.

\textbf{OMat24}: The Open Materials 2024 (OMat24) dataset is a large-scale collection of non-equilibrium structures designed to maximize structural and compositional diversity. It was generated by perturbing relaxed structures from the Alexandria dataset\cite{schmidt2023machine} via three approaches: (i) Rattled Boltzmann Sampling, in which atomic positions and unit cells were randomly perturbed and candidates were selected via a Boltzmann-like distribution at temperatures of 300, 500, and 1000 K; (ii) \textit{ab initio} molecular dynamics, with short (50-step) NVT and NPT trajectories at 1000 and 3000 K; and (iii) Rattled Relaxations, where randomly perturbed Alexandria structures were re-relaxed with DFT. All intermediate structures along each relaxation trajectory were included in the dataset. In total, OMat24 contains $\sim$118 million structures labeled with total energies, atomic forces, and cell stresses.\cite{barroso2024open}

\subsection{Density Functional Theory Calculations}
Density functional theory calculations were performed to establish a high-fidelity benchmark dataset using the Vienna \textit{ab initio} Simulation Package (VASP) with the projector augmented wave (PAW) method.\cite{kresse_efficient_1996,blochl1994projector} We employed the Perdew-Burke-Ernzerhof (PBE) functional,\cite{perdew_generalized_1996} applying effective Hubbard $U$ corrections\cite{dudarev_electron-energy-loss_1998} to $3d$ transition metals. The U values are consistent with those used in the Materials Project\cite{wang2006oxidation, jain2011formation}. The calculations used a plane-wave energy cutoff of 450 eV and a $k$-point density of at least 100 per reciprocal volume. Convergence criteria for the electronic energies and ionic forces were set to $10^{-5}$ eV and 0.05 eV \AA$^{-1}$, respectively.

\subsection{Ion Migration Barriers}
The ion migration energy barriers and minimum energy pathways were calculated using the climbing image nudged elastic band (CI-NEB) method.\cite{henkelman_climbing_2000, henkelman_improved_2000} For each material, all symmetrically distinct Li/Na-ion migration pathways with lengths shorter than 5 \r{A} were examined. The migration energy barrier ($E_a$)\cite{van2001first} is defined as the kinetically resolved activation barrier, which accounts for both forward and backward hops by referencing the average energies of the initial and final states:
\begin{equation}
    E_a = E(\sigma_{t}) - \frac{1}{2}[E(\sigma_{i})+E(\sigma_{f})]
\end{equation}
where $E(\sigma_t)$, $E(\sigma_i)$ and $E(\sigma_t)$ are the energies of the saddle, initial, and final states, respectively.\cite{van2001first}

\subsection{Molecular Dynamics Simulations}
All molecular dynamics (MD) simulations were conducted in the NVT ensemble using a Nose-Hoover thermostat \cite{hoover1985canonical} at temperatures of 800 K and 1200 K. The initial structures were fully relaxed until atomic forces were below 0.05 eV/\r{A}. The simulations were performed for 50 ps with a timestep of 2 fs. The mean square displacement (MSD) of \ce{Li+}/\ce{Na+} was extracted from snapshots of the MD trajectories at each time step. The diffusivity was determined by fitting the MSD data from the last 20 ps of each simulation using linear regression. The ion diffusivity was calculated as follows:
\begin{equation}
       D_{T} = \frac{\left<\Delta \mathrm{r}(t)\right>^2}{2dt}
\end{equation}
where $d$ is the dimensionality (equal to three for bulk diffusion), and $<\Delta r(t)>^2$ is the MSD of \ce{Li+}/\ce{Na+} over a time interval $t$ at temperature $T$. We note that three materials (\ce{Na3SbS4}, \ce{Na3PS4}, \ce{Na3PSe4}) were excluded from the diffusivity analysis due to MSD values below 5 \r{A} in the AIMD simulations at 800 K, indicating insufficient ionic mobility for reliable diffusion coefficient extraction.

The activation energy ($E_a$) was estimated using the Arrhenius relationship, $D = D_{0}e^{\frac{-E_{a}}{kT}}$, where $T$ is the temperature, $k$ is Boltzmann's constant, and $D_0$ is the pre-exponential factor. $E_a$ was derived from the slope of the linear relationship between $\ln(D)$ and $\frac{1}{T}$ using the diffusivity values obtained at 800 K and 1200 K.

The radial distribution functions (RDF)\cite{PhysRevA.7.2130} for each distinct pair of atomic species were computed using the following equation:
\begin{equation}
g(r) = \frac{1}{4\pi r^2}\frac{1}{N\rho}\sum_i^N{\sum_{j\neq i}^N{\delta(r-||x_i-x_j||)}}
\end{equation}
The RDFs were calculated and averaged over frames extracted at 0.2 ps intervals from the 30-50 ps portion of the MD trajectories. The mean absolute errors (MAEs) between uMLIPs and AIMD RDFs were calculated by integrating over the radial distance $r$:
\begin{equation}
\mathrm{MAE_{RDF}} = \int_{0}^{r_{\rm cut}}|\langle g(r)\rangle_{\rm AIMD} - \langle g(r)\rangle_{\rm MLIP}|dr
\end{equation}
where $\langle g(r)\rangle_{\rm AIMD}$ and $\langle g(r)\rangle_{\rm MLIP}$ are the averaged equilibrium RDF from AIMD and uMLIP simulations, respectively. The cutoff radius $r_{cut}$ was set to 10 \r{A}.

\section{Data availability}
All data generated and analyzed in this study are provided in the Figshare repository at https://figshare.com/s/9ae75373f32ddfe13d2d.

\section{Acknowledgements}
X. Guo acknowledges the financial support of the The National Natural Science Foundation of China (NSFC) Young Scientists Fund-C(22509173), Z. Wang acknowledges the support of the City University of Hong Kong Grant (9229162). Some calculations were performed using the computational facilities of CityU Burgundy, which are provided and managed by the Computing Services Centre at the City University of Hong Kong.

\section{Author contributions}
X.G. conceived the project. X.G. generated the DFT reference data, conducted uMLIPs calculations, and performed the data analysis. X.G. wrote the original draft. C.G conducted uMLIP calculations. Z.W. performed data analysis and edited the manuscript. All authors read, edited, and approved the ﬁnal manuscript.

\section{Competing Interests}
The authors declare no competing interests.

\bibliography{refs.bib}

\end{document}


\maketitle

\clearpage


\begin{figure}[htp!]
\centering
{\includegraphics[width=\textwidth]{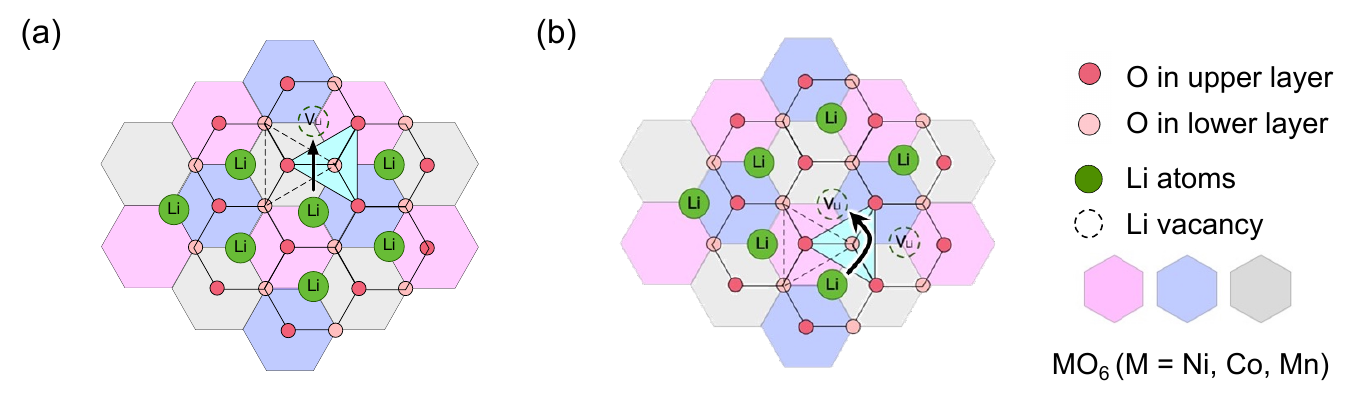}}
\caption{Schematic diagrams of (a) single-vacancy and (b) di-vacancy migration mechanisms in layered cathode materials.}
\label{fig:barrier_compare} 
\end{figure}

\clearpage


\begin{table}[htbp]
\renewcommand{\arraystretch}{1.2}
    \centering
    \begin{tabular}{l|l|c}
    \toprule
    
    Model identifier & Notation & Ref. \\
    \midrule
         M3GNet-MP-2021.2.8-PES &  M3GNet (MPF) & \citenum{chen2022universal} \\
         M3GNet-MP-2021.2.8-DIRECT-PES &  M3GNet-DIRECT (MPF) & \citenum{qi2024robust} \\
         CHGNet-MPtrj-2024.2.13-11M-PES &  CHGNet (MPTrj) & \citenum{deng2023chgnet} \\
         MACE-MP-0b3 &  MACE (MPTrj) & \citenum{Batatia2022Design, Batatia2022mace} \\
         MACE-MPA-0 &  MACE (MPA) & \citenum{Batatia2022Design, Batatia2022mace} \\
         MACE-OMAT-0 &  MACE (OMat24) & \citenum{Batatia2022Design, Batatia2022mace} \\
         orb-mptraj-only-v2 &  Orb-v2 (MPTrj) & \citenum{neumann2024orb} \\
         orb-v2 &  Orb-v2 (MPA) & \citenum{neumann2024orb} \\
         orb-v3-conservative-20-mpa &  Orb-v3 conservative 20 (MPA) & \citenum{rhodes2025orbv3}\\
         orb-v3-conservative-inf-mpa &  Orb-v3 conservative inf (MPA) & \citenum{rhodes2025orbv3}\\
         orb-v3-conservative-20-omat &  Orb-v3 conservative 20 (OMat24) & \citenum{rhodes2025orbv3}\\
         and orb-v3-conservative-inf-omat &  Orb-v3 conservative inf (OMat24) & \citenum{rhodes2025orbv3}\\
         orb-v3-direct-20-mpa &  Orb-v3 direct 20 (MPA) & \citenum{rhodes2025orbv3}\\
         orb-v3-direct-inf-mpa &  Orb-v3 direct inf (MPA) & \citenum{rhodes2025orbv3}\\
         orb-v3-direct-20-omat &  Orb-v3 direct 20 (OMat24) & \citenum{rhodes2025orbv3}\\
         orb-v3-direct-inf-omat &  Orb-v3 direct inf (OMat24) & \citenum{rhodes2025orbv3}\\
         mattersim-v1.0.0-5M &  MatterSim (MPA active learning) & \citenum{yang2024mattersim} \\
         GRACE-2L-MP-r6 &  GRACE (MPTrj) & \citenum{lysogorskiy2025graph}\\
         GRACE-2L-OMAT-L-base &  GRACE (OMat24) & \citenum{lysogorskiy2025graph}\\
         GRACE-2L-OAM-L &  GRACE fine-tune (OMat24 + MPA) & \citenum{lysogorskiy2025graph}\\
         SevenNet-0 &  SevenNet (MPTrj) & \citenum{park_scalable_2024}\\
         SevenNet-omat &  SevenNet (OMat24) & \citenum{park_scalable_2024} \\
         SevenNet-MF-ompa &  SevenNet multi-fidelity (MPA + OMat24) & \citenum{kim_sevennet_mf_2024}\\
    \bottomrule
    \end{tabular}
    \caption{Nomenclature of benchmarked uMLIP models. The table maps the original model identifiers to the simplified notation used throughout this study, along with their respective references.}
    \label{tab:notations}
\end{table}

\clearpage


\begin{figure}[htp!]
\centering
{\includegraphics[width=\textwidth]{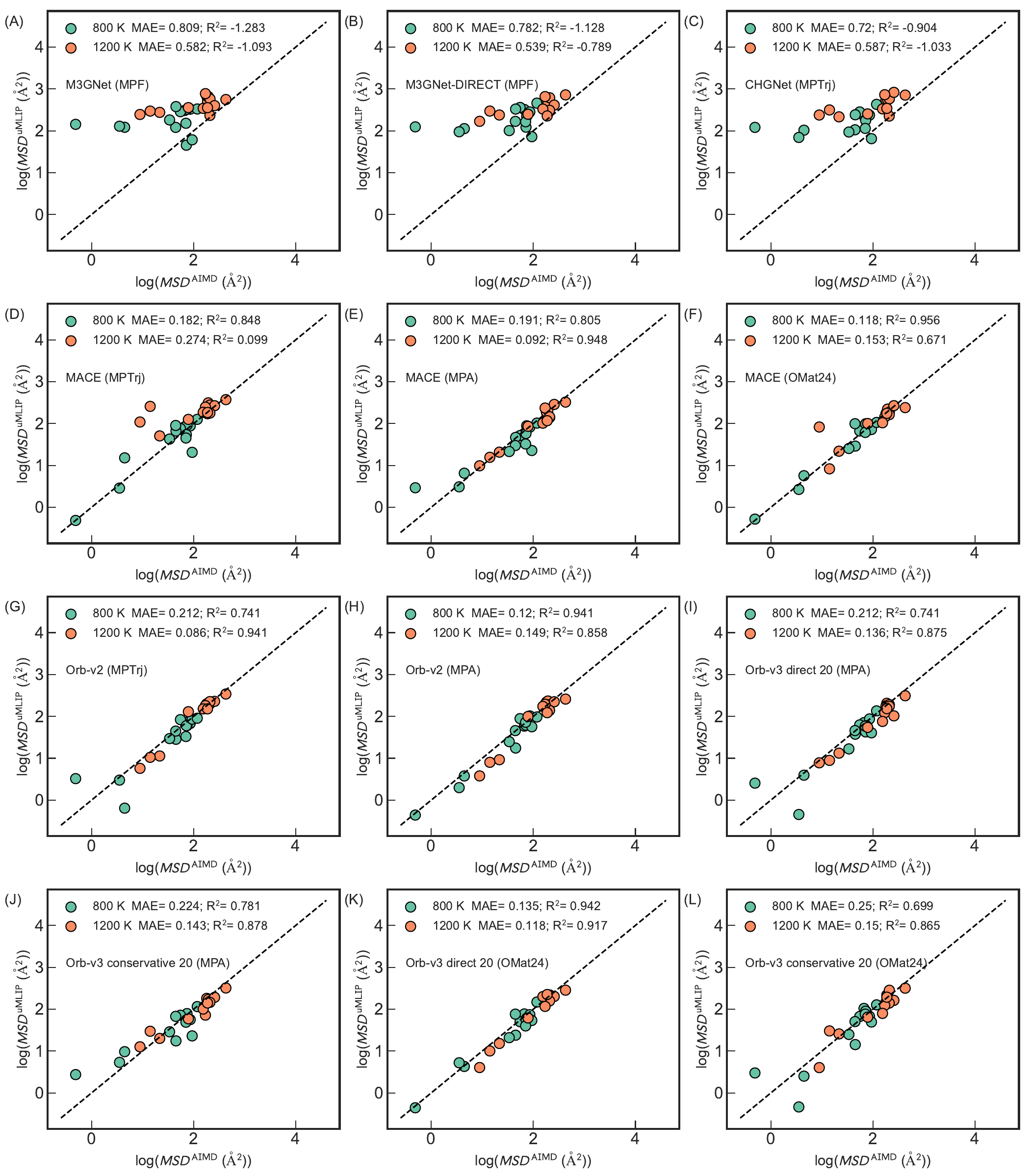}}
\end{figure}

\clearpage

\begin{figure}[htp!]
\centering
{\includegraphics[width=\textwidth]{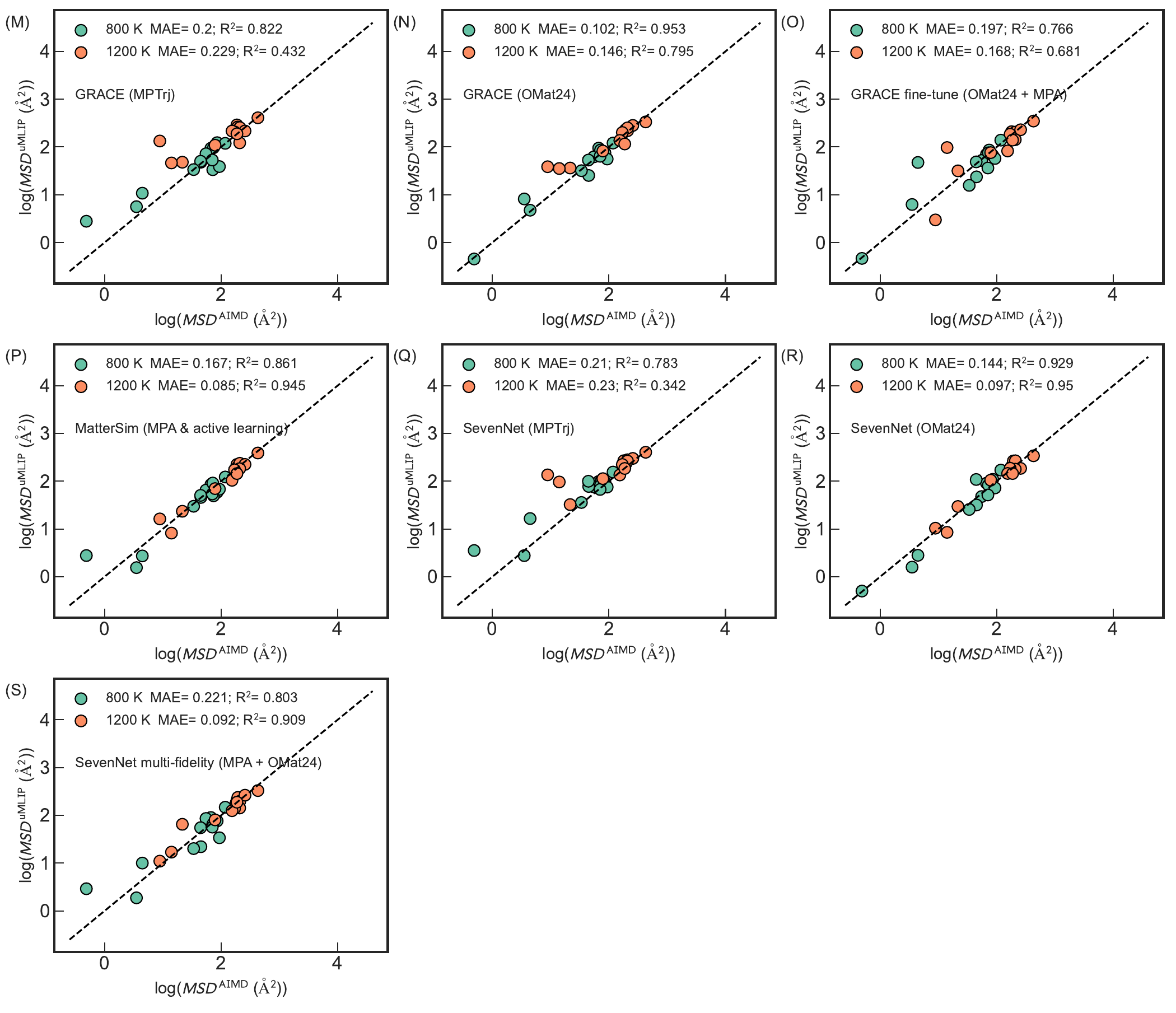}}
\caption{Comparison of log(MSD) obtained from AIMD and uMLIPs simulations over 50 ps NVT molecular dynamics at 800 K and 1200 K.}
\label{fig:fig_s1} 
\end{figure}

\clearpage


\begin{figure}[htp!]
\centering
{\includegraphics[width=\textwidth]{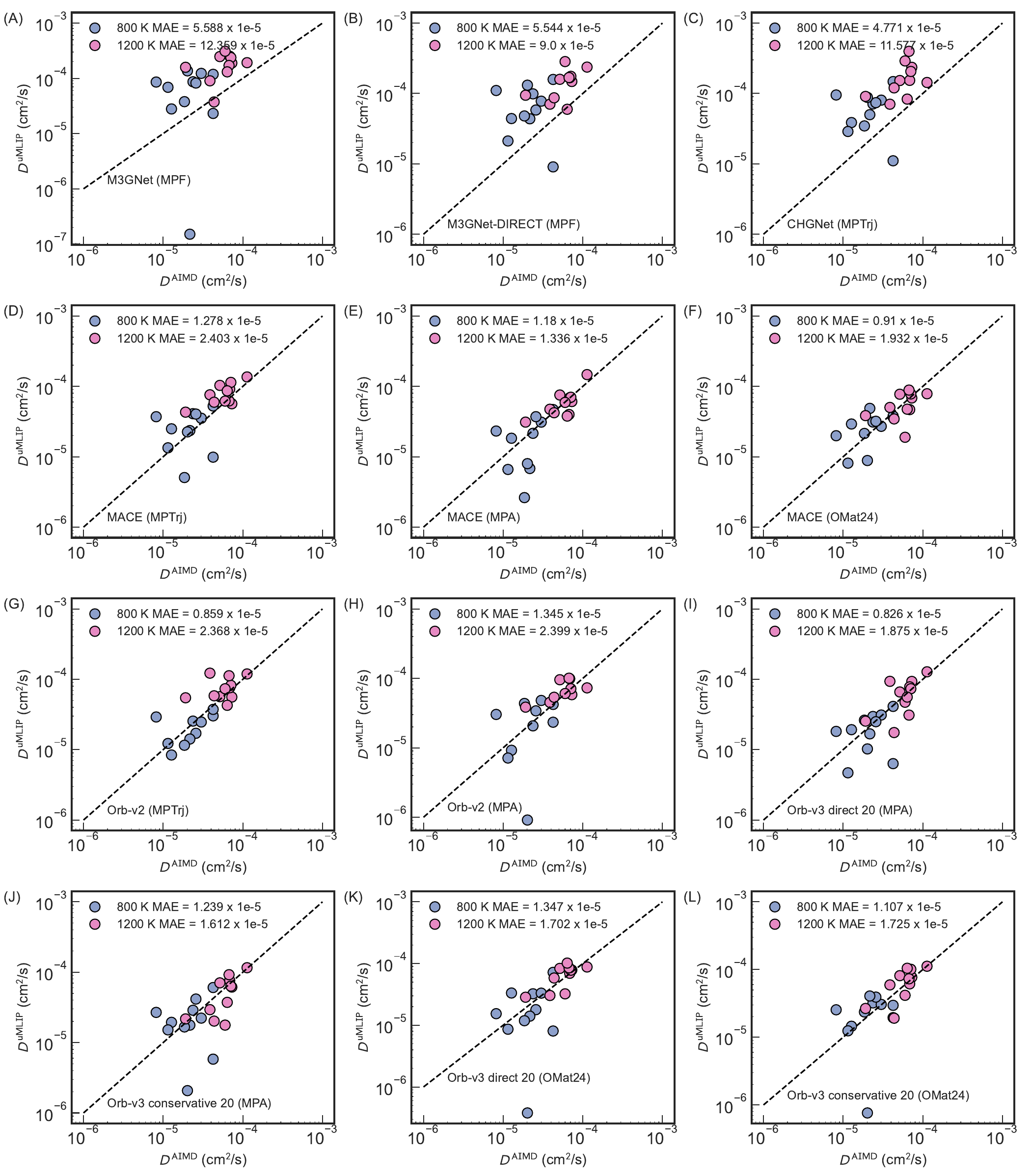}}
\end{figure}

\clearpage

\begin{figure}[htp!]
\centering
{\includegraphics[width=\textwidth]{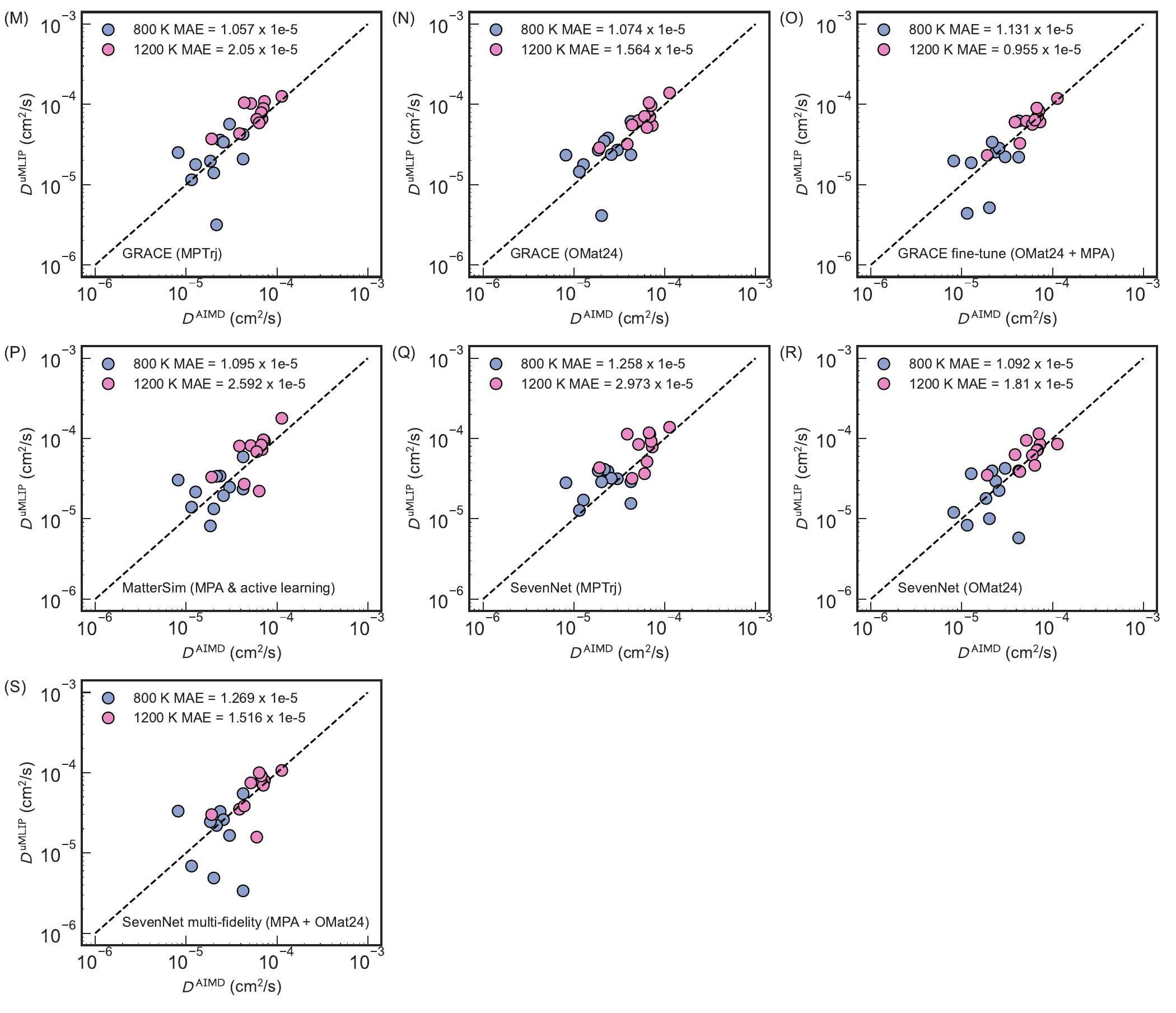}}
\caption{Comparison of \ce{Li+}/\ce{Na+} diffusivity obtained by AIMD and uMLIPs simulations based on 50 ps NVT molecular dynamics at 800 K and 1200 K. Several calculations yield unphysical negative diffusivity values; these cases are listed in Table S2.}
\label{fig:fig_s2} 
\end{figure}

\begin{table}[htbp]
\renewcommand{\arraystretch}{1.5}
\centering
\begin{tabular}{llcc}
\toprule
uMLIP & Material & Temperature (K) & Diffusivity (cm$^{2}$/s) \\
\midrule
MACE (MPA) & \ce{Li3YCl6} & 800  & $-1.153\times 10^{-7}$ \\
Orb-v2 (MPTrj) & \ce{Li6PS5Cl} & 800 & $-9.10\times 10^{-7}$ \\
Orb-v2 (MPA) & \ce{Li3YBr6} & 800 & $-3.54\times 10^{-6}$ \\
Orb-v2 (MPA) & \ce{Na10SiP2S12} & 1200 & $-1.28\times 10^{-6}$ \\
GRACE finetune (OMat24+MPA) & \ce{Na10GeP2S12} & 800 & $-1.56\times 10^{-6}$ \\
SevenNet multi-fidelity (OMat24+MPA) & \ce{Na10SiP2S12} & 800 & $-1.09\times 10^{-6}$ \\
\bottomrule
\end{tabular}
\caption{Cases exhibiting unphysical negative diffusivity values in 50 ps NVT molecular dynamics simulations, together with the corresponding uMLIP model, material, and temperature.}
\label{tab:outlier_diffusivity}
\end{table}

\clearpage


\begin{figure}[htp!]
\centering
{\includegraphics[width=\textwidth]{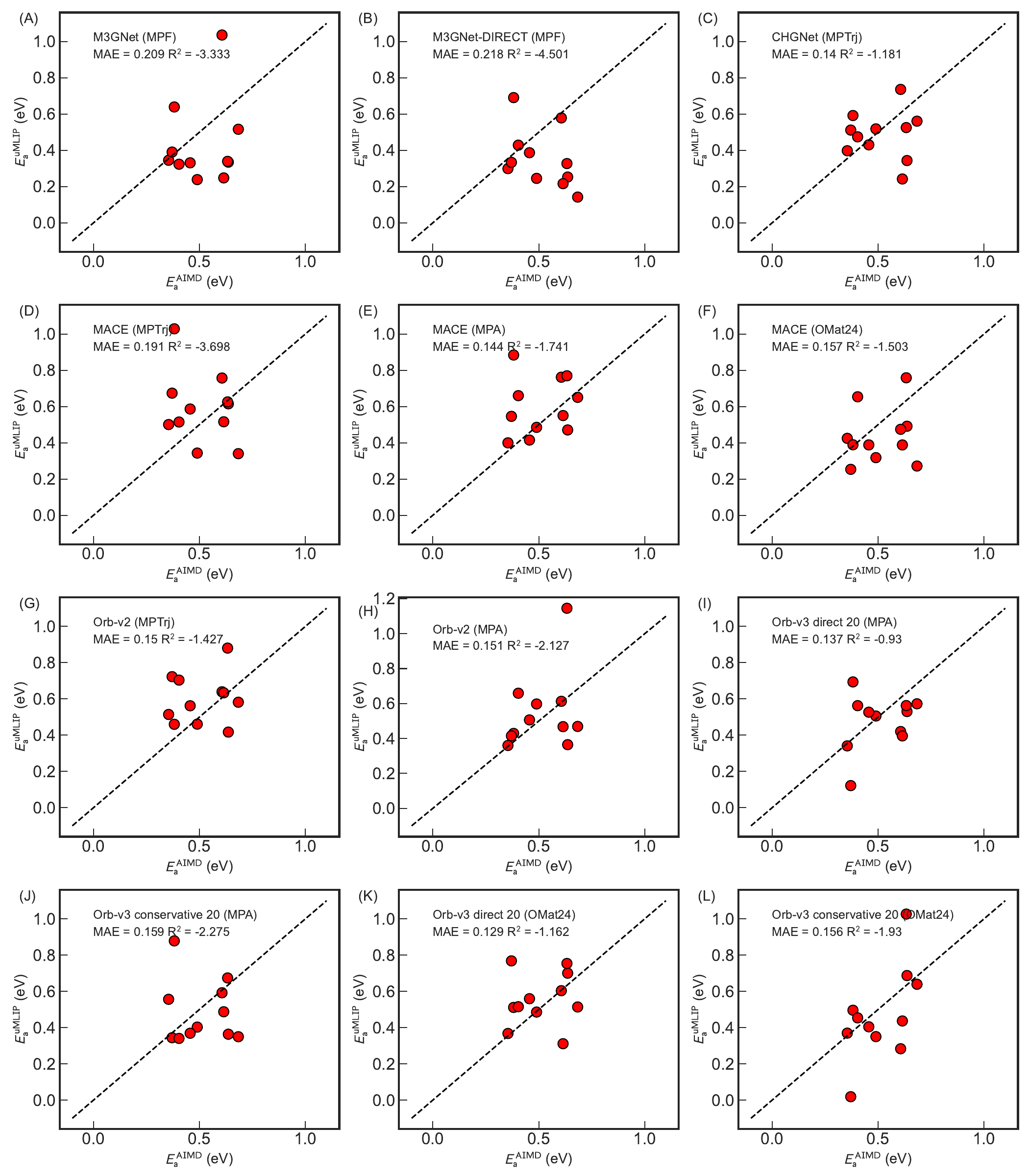}}
\end{figure}

\clearpage

\begin{figure}[htp!]
\centering
{\includegraphics[width=\textwidth]{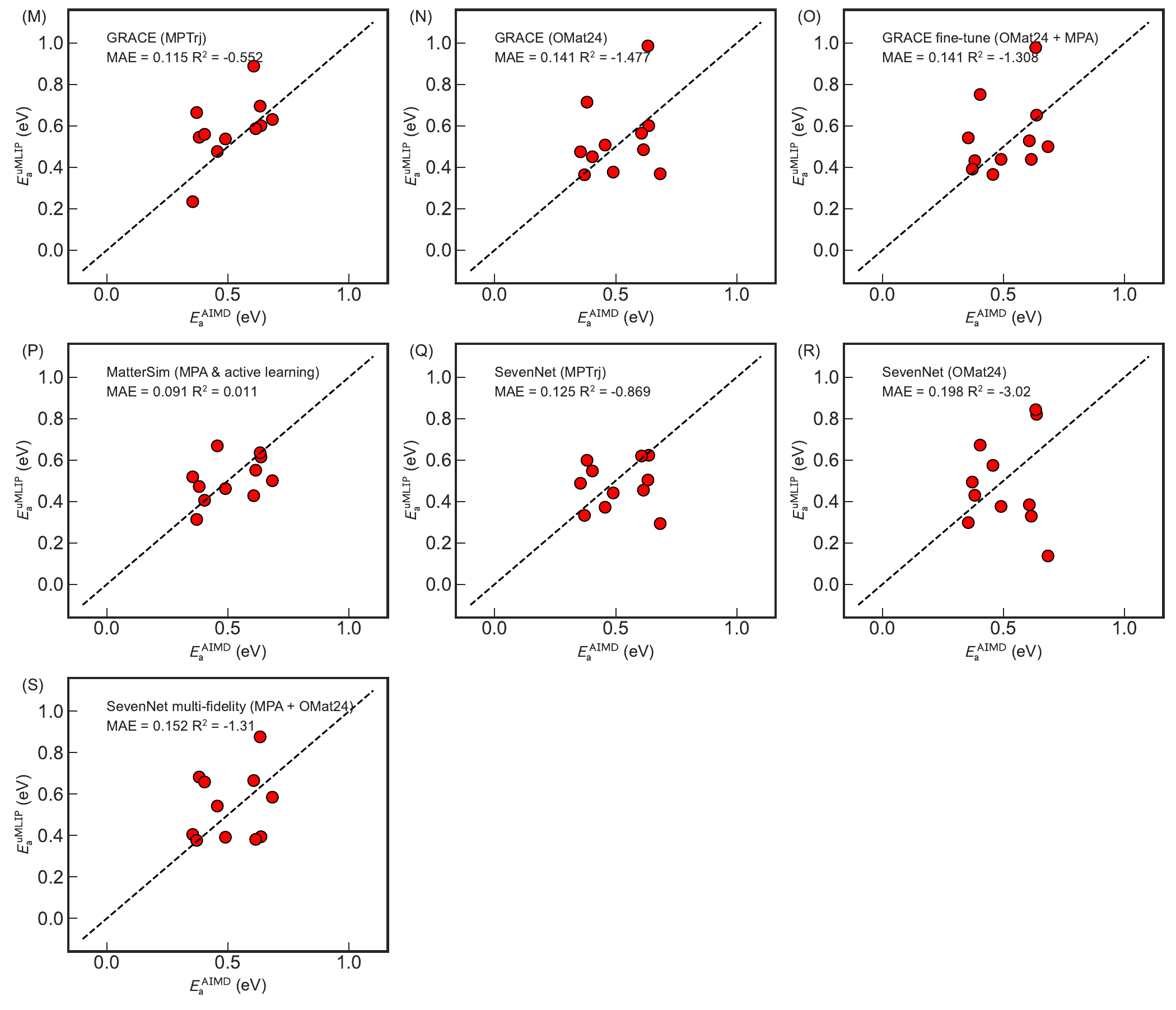}}
\caption{Comparison of \ce{Li+}/\ce{Na+} activation energies derived from AIMD and uMLIPs. Results are calculated from 50 ps NVT molecular dynamics simulations.}
\label{fig:fig_s3} 
\end{figure}

\clearpage


\begin{figure}[htbp]
    \centering
    \includegraphics[width=\textwidth]{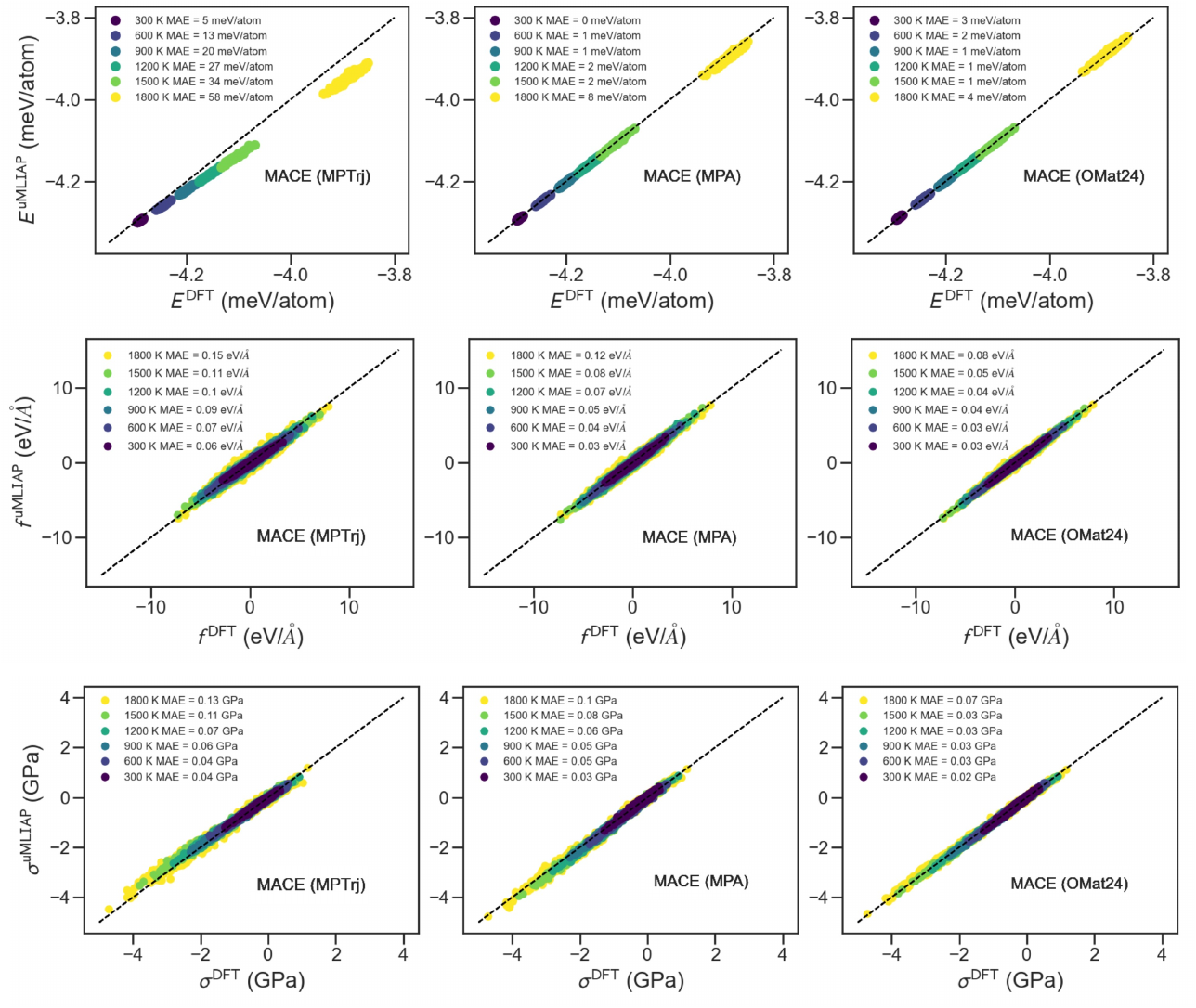}
    \caption{Comparison of (a) energies, (b) forces, and (c) stresses obtained from DFT and MACE models trained on various datasets, using snapshots from ab initio molecular dynamics (AIMD) simulations conducted at temperatures ranging from 300 K to 1800 K.}
    \label{fig:fig_s4}
\end{figure}

\clearpage


\begin{figure}[htbp]
    \centering
    
    \includegraphics[width=\textwidth]{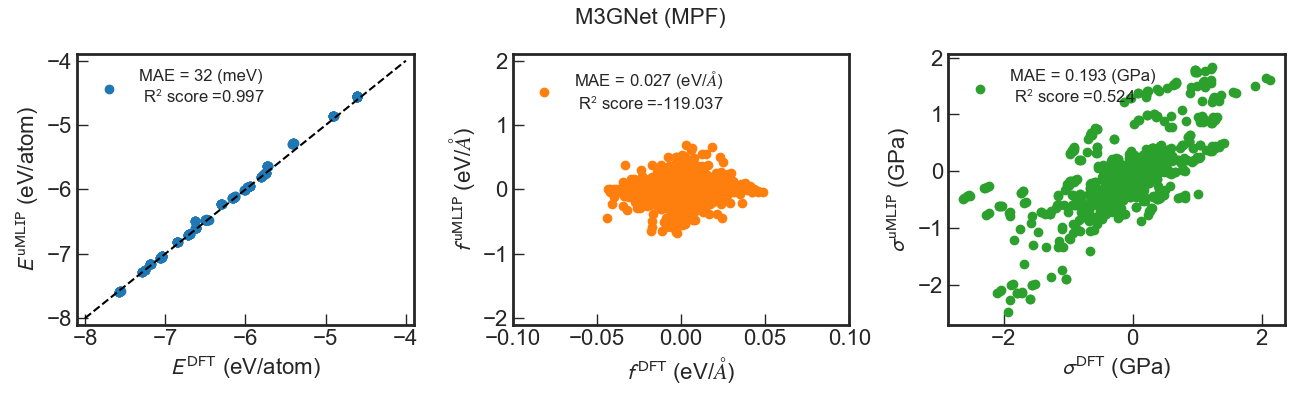}
    \includegraphics[width=\textwidth]{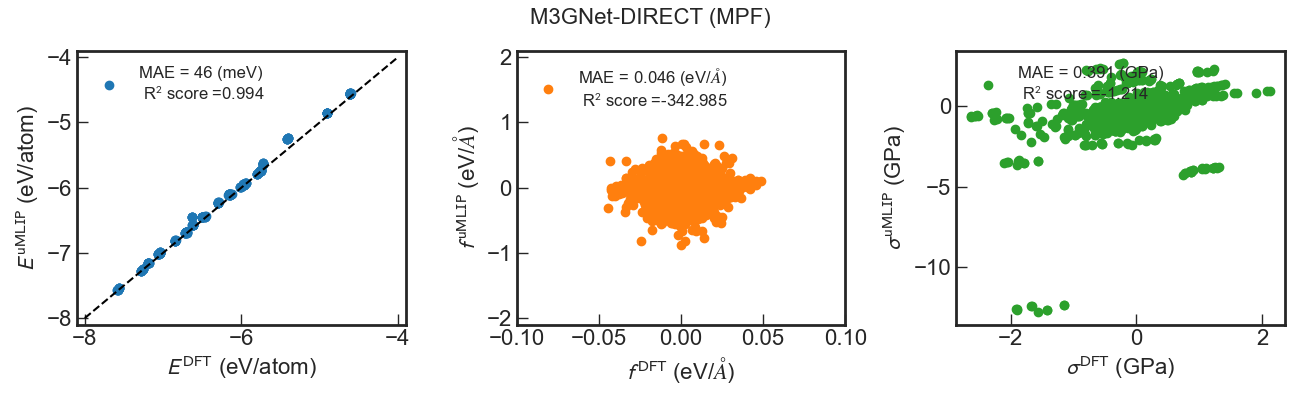}
    \includegraphics[width=\textwidth]{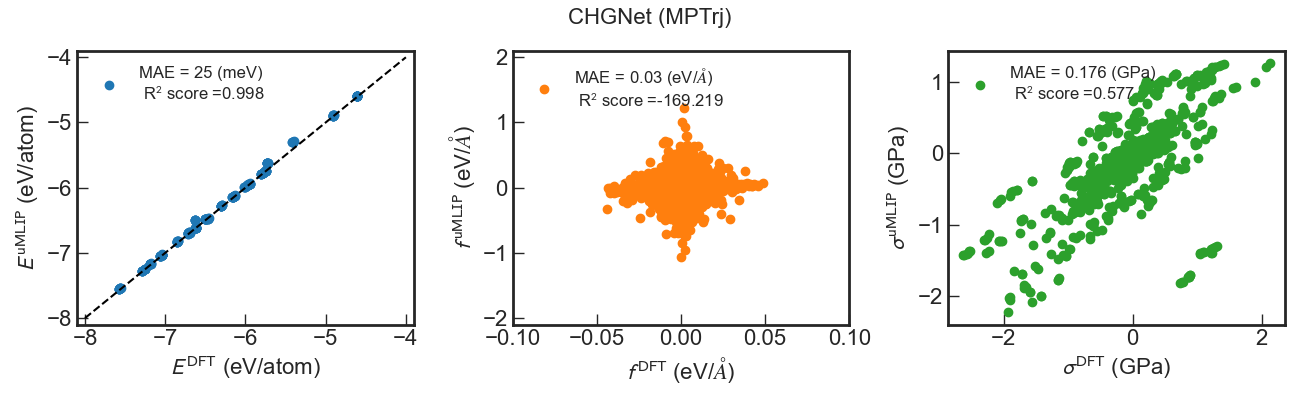}
\end{figure}

\begin{figure}[htbp]
    \centering

    \includegraphics[width=\textwidth]{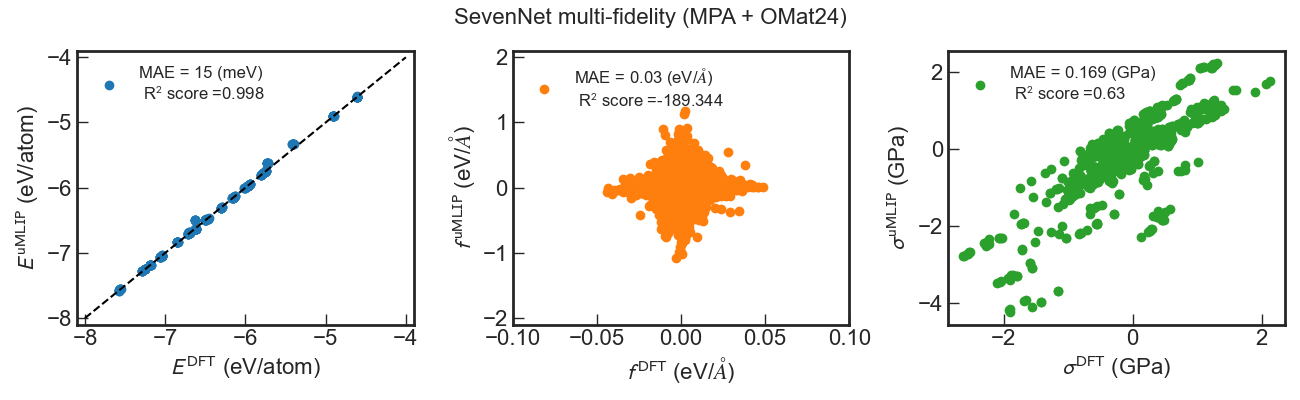}
    \includegraphics[width=\textwidth]{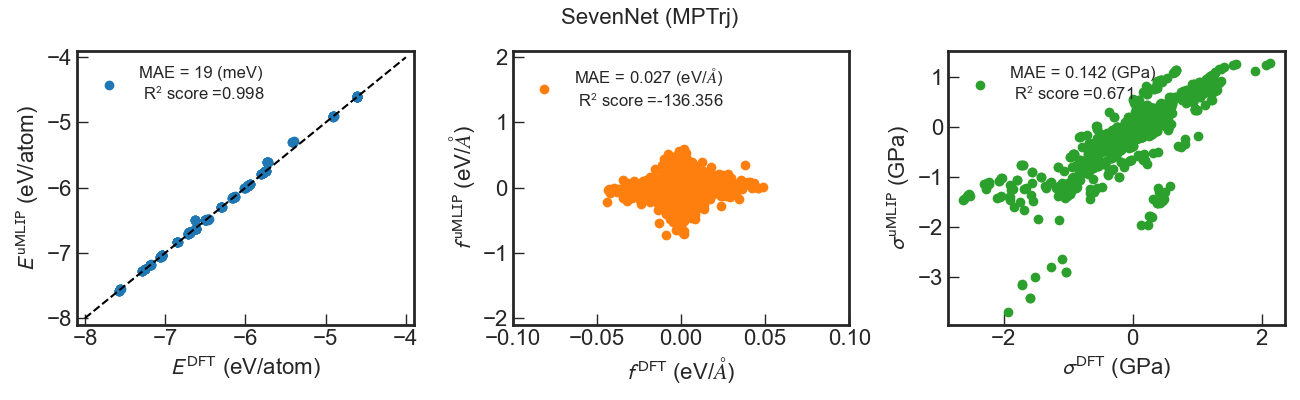}
    \includegraphics[width=\textwidth]{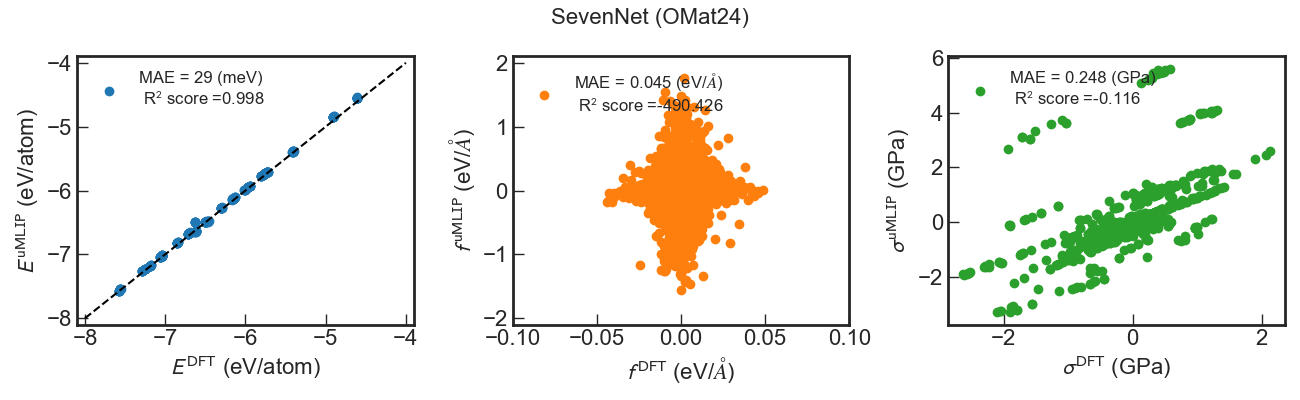}

\end{figure}

\begin{figure}[htbp]
    \centering
    
    \includegraphics[width=\textwidth]{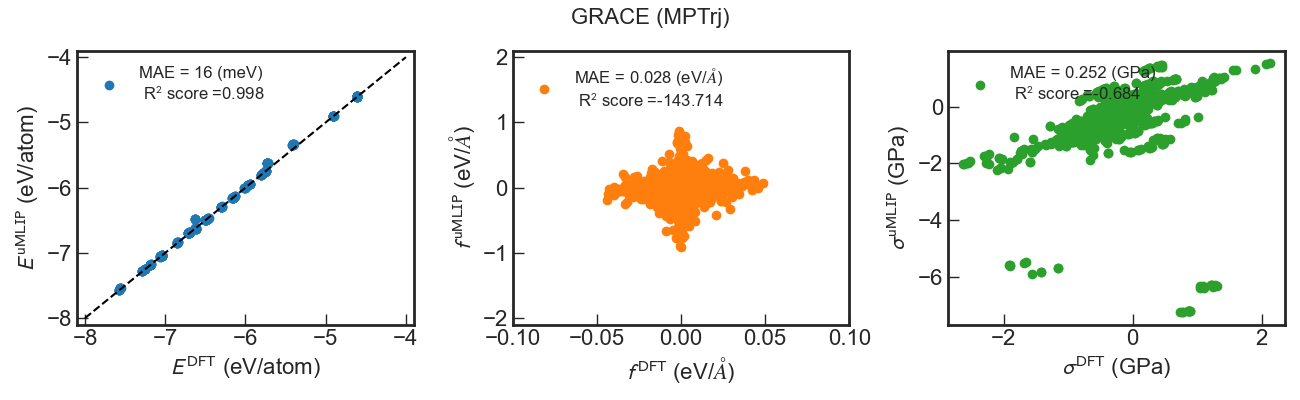}
    \includegraphics[width=\textwidth]{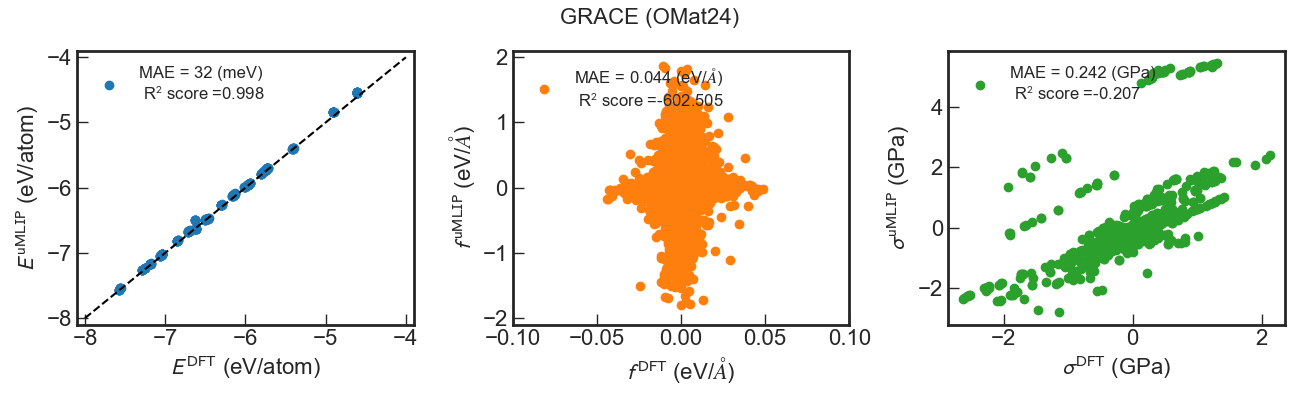}
    \includegraphics[width=\textwidth]{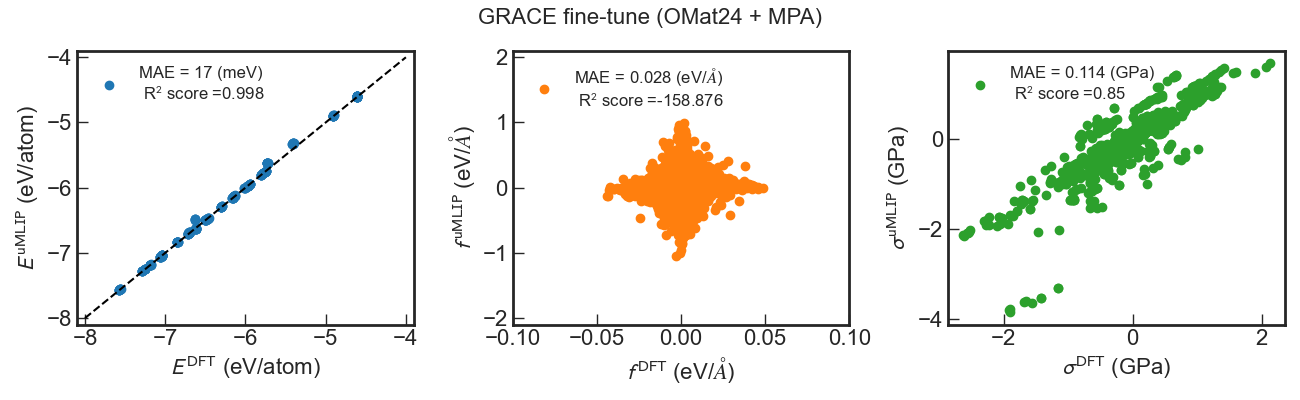}

\end{figure}

\begin{figure}[htbp]
    \centering
    
    \includegraphics[width=\textwidth]{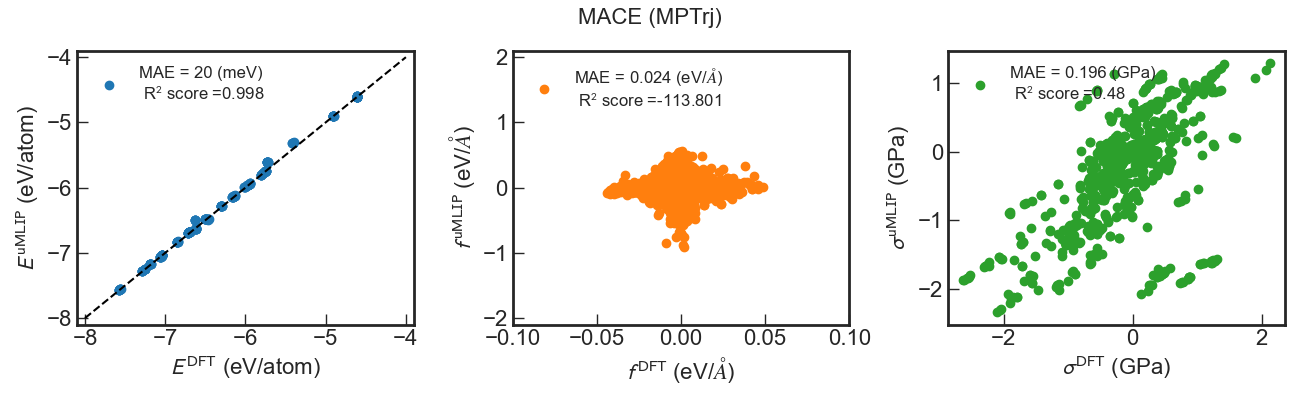}
    \includegraphics[width=\textwidth]{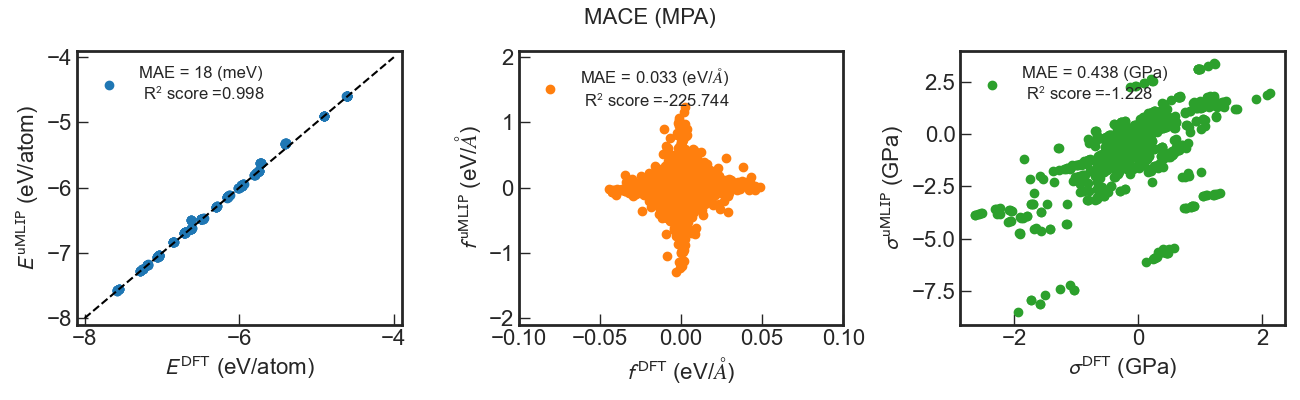}
    \includegraphics[width=\textwidth]{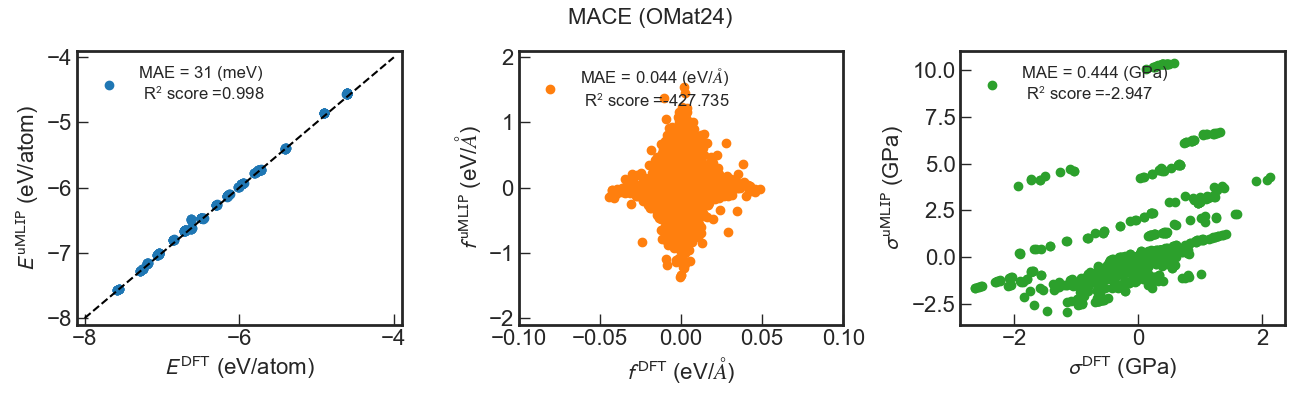}
\end{figure}

\begin{figure}[htbp]
    \centering
    \includegraphics[width=\textwidth]{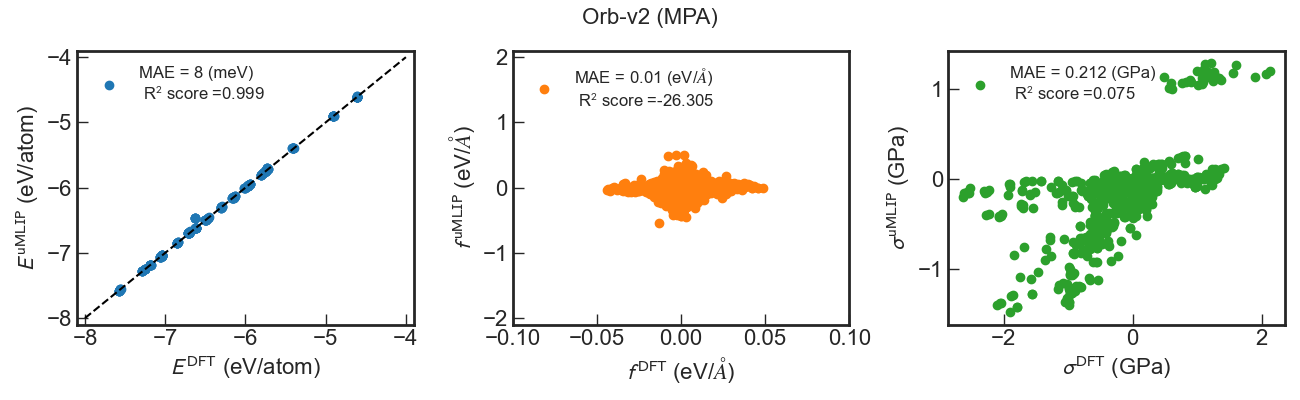}
    \includegraphics[width=\textwidth]{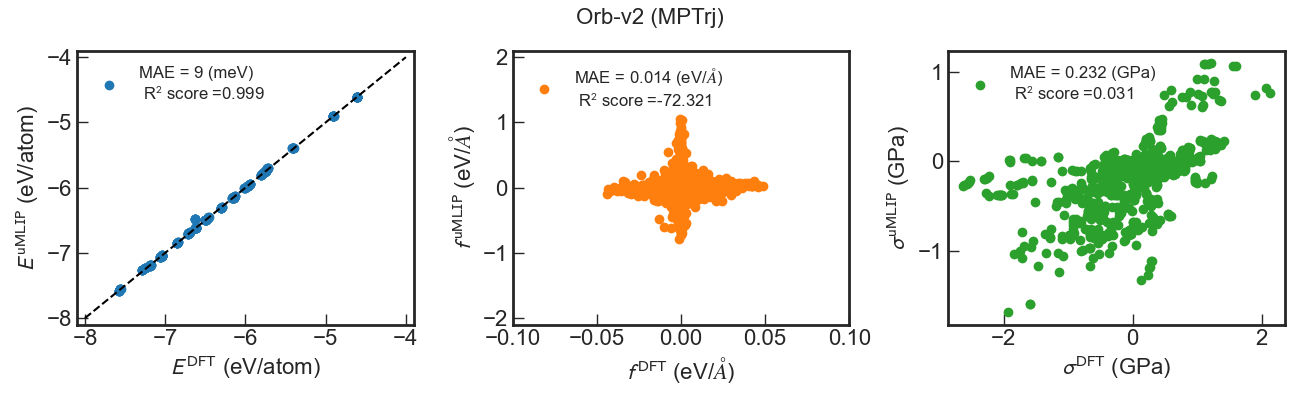}

\end{figure}

\begin{figure}[htbp]
    \centering
    \includegraphics[width=\textwidth]{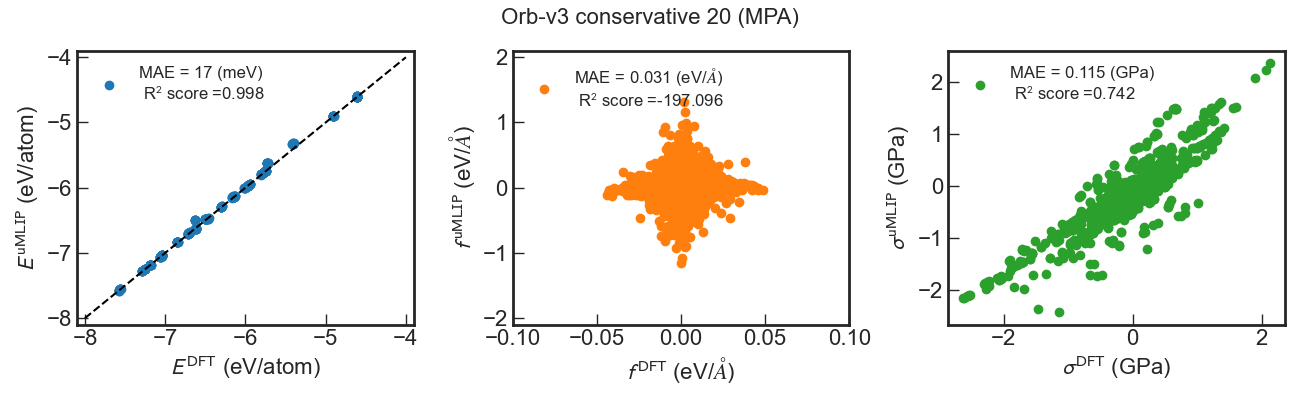}
    \includegraphics[width=\textwidth]{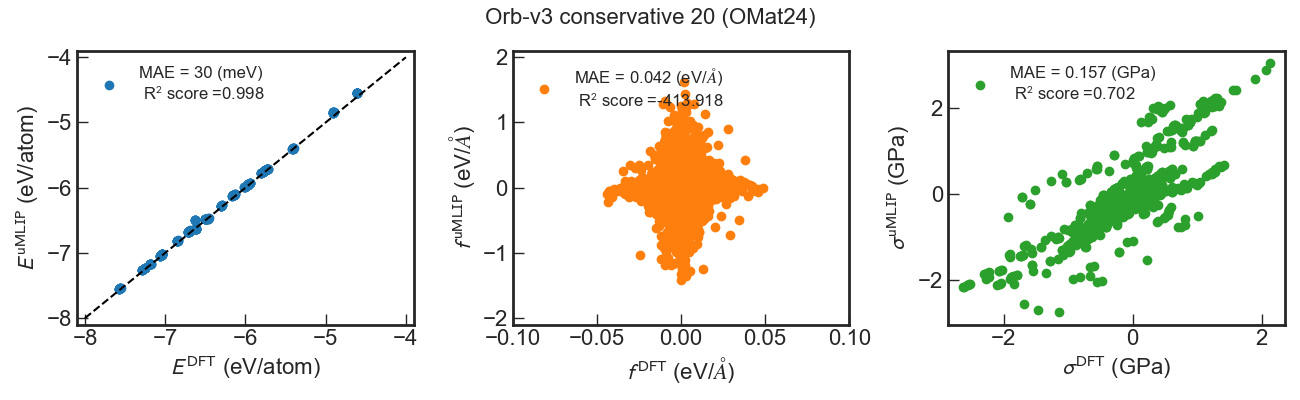}
    \includegraphics[width=\textwidth]{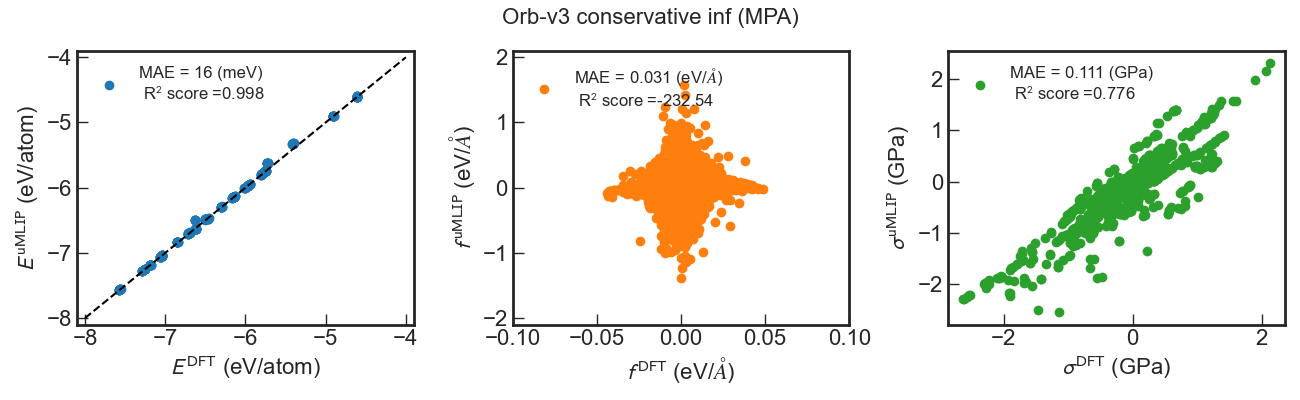}
    \includegraphics[width=\textwidth]{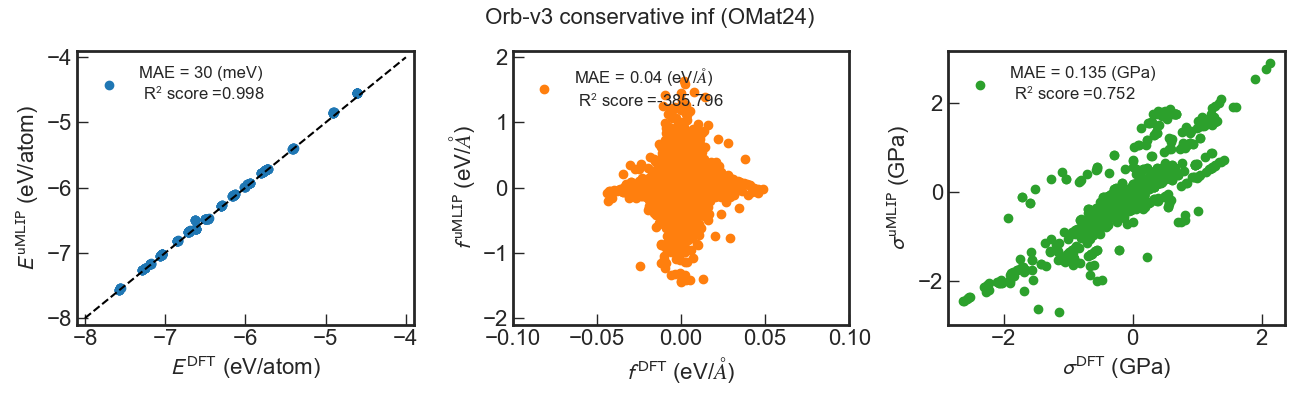}

\end{figure}

\begin{figure}[htbp]
    \centering
    
    \includegraphics[width=\textwidth]{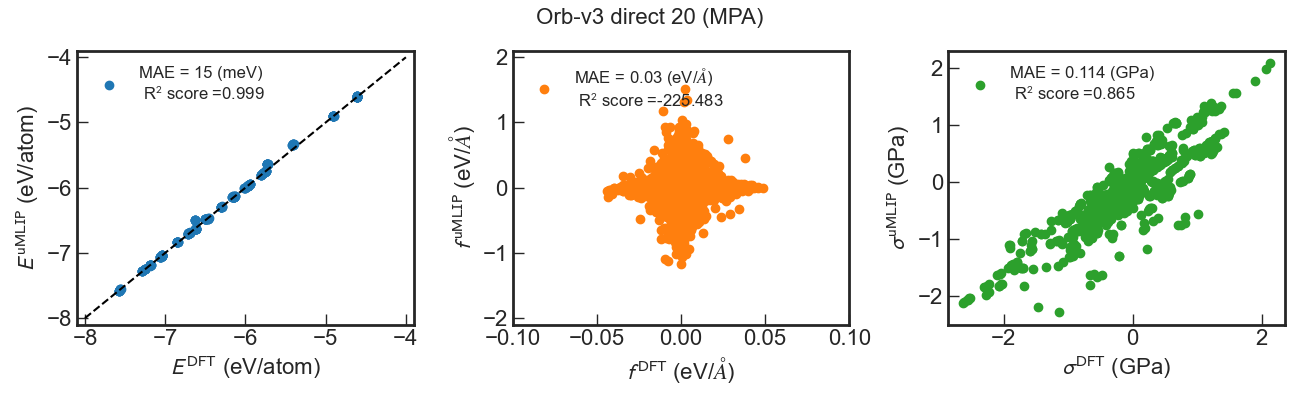}
    \includegraphics[width=\textwidth]{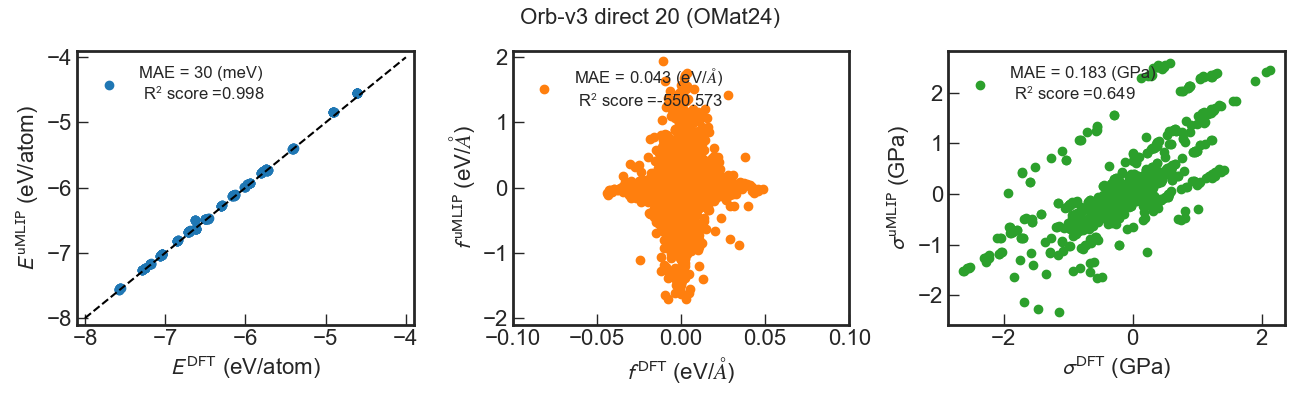}
    \includegraphics[width=\textwidth]{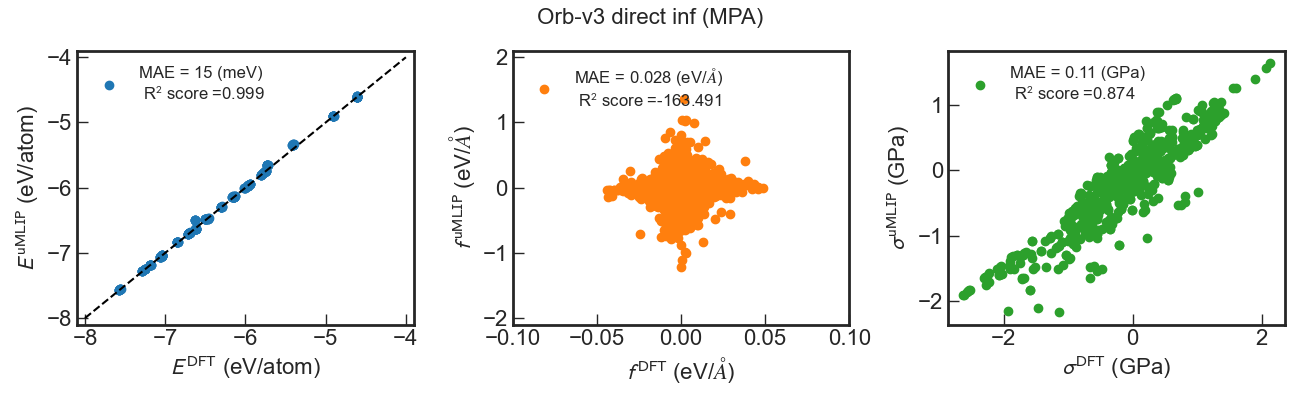}
    \includegraphics[width=\textwidth]{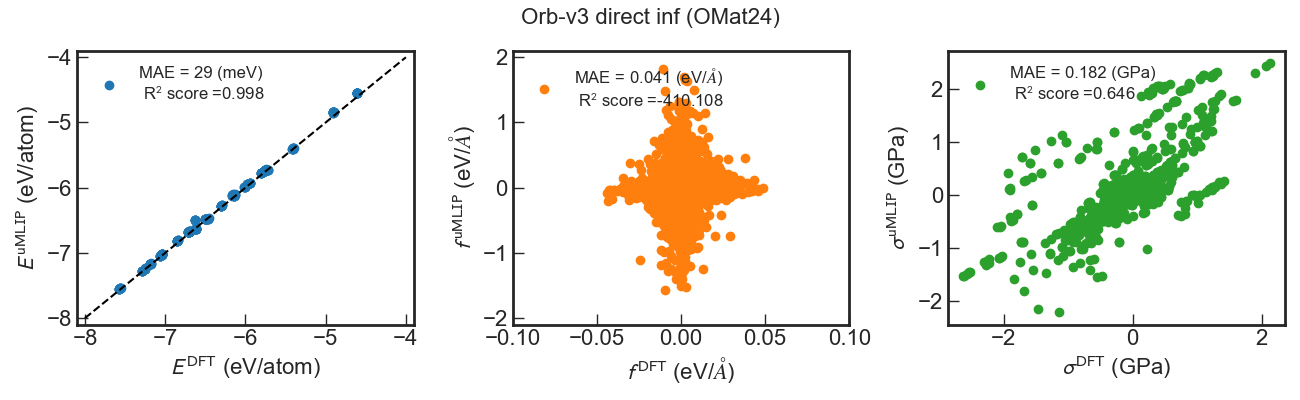} 
\end{figure}

\clearpage

\begin{figure}[htbp]
    \centering
\includegraphics[width=\textwidth]{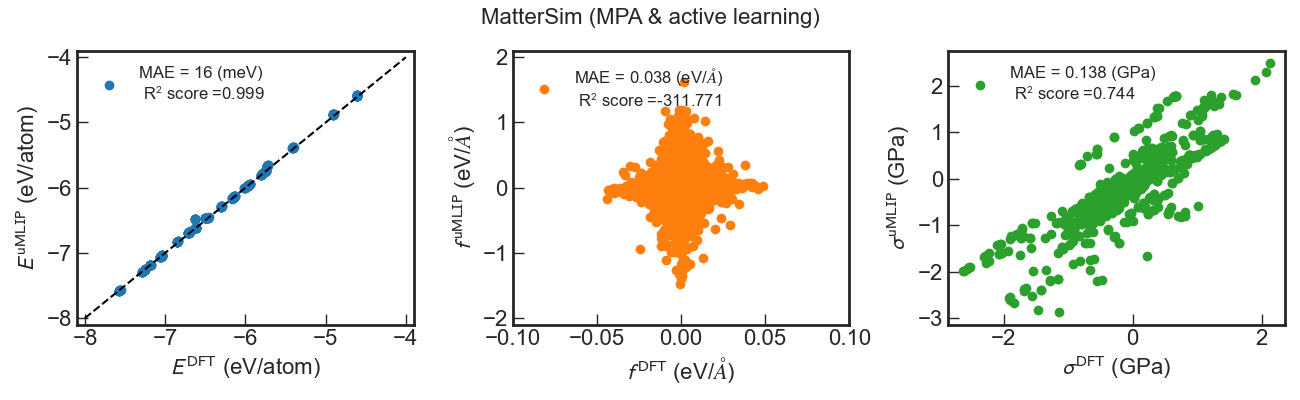}
\caption{Comparison of energies, forces normal to the ion‑migration pathway, and stresses of the transition states obtained from DFT and various uMLIP models. The configuration of the transition states were obtained through well-converged NEB calculations using DFT.}
    \label{fig:fig_s5}
\end{figure}

\clearpage

\bibliography{refs.bib}